\documentclass[12p]{article}
\usepackage{latexsym}
\usepackage{color}
\usepackage{amsmath}
\usepackage{epsfig}
\usepackage{hyperref}
 \usepackage{subfigure}
 \usepackage{dcolumn}% Align table columns on decimal point
   \usepackage{threeparttable}

\newcommand{\ortala}[1]{\begin{center}#1\end{center}}

\newcommand{\sandd}[1]{\left\langle #1\right\rangle}

\newcommand{\integ}[3]{{{\underset{#1 }{\overset{#2}{\displaystyle\int}}}#3}}

\newcommand{\summ}[3]{{{\underset{#1 }{\overset{#2}{\displaystyle\sum}}}#3}}

\newcommand{\re}[1]{(\ref{#1})}

\newcommand{\eq}[2]{\begin{equation}\label{#1}  #2\end{equation}}

\newcommand{\paran}[1]{\left(#1\right)}

\newcommand{\sch}[1]{Schrodinger}

\newcommand{\komb}[2]{\paran{\begin{array}{c} #1 \\ #2 \end{array}}}

\usepackage{amssymb}
\usepackage{latexsym}
\begin{document}

\ortala{\large\textbf{Random Field Distributed Heisenberg Model on a Thin Film Geometry}}

\ortala{\textbf{\"Umit Ak\i nc\i \footnote{\textbf{umit.akinci@deu.edu.tr}}}}

\ortala{\textit{Department of Physics, Dokuz Eyl\"ul University,
TR-35160 Izmir, Turkey}}

\section{Abstract}
The effects of the bimodal random field distribution on the thermal
and magnetic properties of  of the  Heisenberg thin film have
been investigated by making use of a two spin cluster
with the decoupling approximation. Particular attention has been
devoted to the obtaining of phase diagrams and magnetization behaviors.
The physical behaviors of special as well as tricritical points
are discussed  for a wide range of selected Hamiltonian parameters.
For example, it is found that  when the strength of magnetic field
increases, the locations of special point in related plane decrease.
Moreover, tricritical behavior has been obtained for the
higher values of the magnetic field, and influences of the varying
Hamiltonian parameters on its behavior have been elucidated
in detail in order to have a better understanding of
mechanism underlying of the considered system.

\section{Introduction}\label{introduction}
Recently, there has been growing interest both theoretically and
experimentally on the finite magnetic materials especially on
semi-infinite systems and thin films. The magnetic properties of
the materials in the presence of the free surfaces are drastically
different from the bulk counterparts. This is because of the fact that,
free surface breaks the translational symmetry, i.e. surface atoms are
embedded in an environment of lower symmetry than that of the
inner atoms \cite{ref1,ref2}. If the strength of the surface
exchange interaction is greater than a critical value, the surface
region can exhibit an ordered phase even if the bulk is paramagnetic
and it has a higher transition temperature  than the bulk one.
The aforementioned situation has been observed
experimentally in Refs. \cite{ref3,ref4,ref5}.
A rigorous review about the surface magnetism
can be found in Ref. \cite{ref6}.

In a thin film geometry, it was experimentally found that,
the Curie temperature and the average magnetic moment per atom
increases with the increasing thickness of the film \cite{ref7,ref8}.
Thickness dependent of Curie temperature  has been also measured in Co \cite{ref9},
Fe \cite{ref10} and Ni \cite{ref11} films. One class of the films which
exhibits a strong uniaxial anisotropy \cite{ref12} can be
modeled by Ising model. These systems have been widely studied in
literature by means of several theoretical methods such
as Monte Carlo (MC) simulations \cite{ref13},
mean field approximation (MFA) \cite{ref14} and effective field
theory (EFT) \cite{ref15}. Indeed Ising thin films keep wide space in
the literature (e.g. see references in Ref. \cite{ref16}). Thin films
which do not exhibit a strong uniaxial anisotropy requires to solve
the Heisenberg model in the thin film geometry.
But in contrast to the Ising counterpart, Heisenberg model in a thin film
geometry has been solved in a limited case. Heisenberg model on a thin film
geometry with Green function method \cite{ref17,ref18,ref19},
renormalization group technique \cite{ref20}, MFA \cite{ref21},
EFT \cite{ref22,ref23} and MC \cite{ref24,ref25}, are among them.
Besides, critical and thermodynamic properties of the
bilayer \cite{ref26,ref27} and multilayer \cite{ref28}
systems have been investigated within the cluster variational
method in the pair approximation.

Working on the random field distributed magnetic systems are important.
Although it is difficult to realize these systems experimentally, certain mappings between
these systems and some other systems make these models valuable.
Most obvious one is, similarity between the diluted antiferromagnets in a
homogenous magnetic field and ferromagnetic systems in the presence of random fields \cite{ref29,ref30}.
Besides, a rich class of experimentally accessible disordered systems
can be described by the random field Ising model (RFIM) such as structural
phase transitions in random alloys, commensurate charge-density-wave systems with
impurity pinning, binary fluid mixtures in random porous media, and the melting of intercalates in layered
compounds such as $TiS_2$ \cite{ref31}.  Also, RFIM has been applied in order to
describe critical surface behavior of amorphous semi-infinite systems \cite{ref32,ref33}
and the magnetization process of garnet film \cite{ref34}. Because of these motivations, Ising model in a
quenched random field has been studied over three decades. The model which is
actually based on the local fields acting on the lattice sites which are taken to be
random according to a given probability distribution, was introduced for the first time by Larkin \cite{ref35}
for superconductors and later generalized by Imry and Ma \cite{ref36}.

On the other hand, there have been less attention paid on the
random field effects on the Heisenberg model. Spin-1/2 isotropic classical
Heisenberg model with bimodal random magnetic field distribution is studied within the EFT
for two spin cluster (which is  abbreviated as EFT-2)  \cite{ref37,ref38} and
within the EFT with probability distribution technique \cite{ref40} has been studied.
Similar results have been obtained such as tricritical behavior. Besides, amorphization
effect for the bimodal random magnetic field distributed isotropic Heisenberg model
has been studied \cite{ref39}. Recently, spin-1/2 anisotropic quantum Heisenberg model
with trimodal random magnetic field distribution is investigated within the EFT-2 \cite{ref41}.
All of these works are related to the bulk systems. Thus,  some questions are open for the Heisenbeg
model in a thin film geometry such as,   whether tricritical behavior exist or not
and the behavior of the special point with the random field distribution.

Thus, the aim of this work is to determine the effect of
the bimodal random magnetic field distribution  on the phase diagrams and
magnetization behavior of the  isotropic Heisenberg thin film. For this aim,
the paper is organized as follows: In Sec. \ref{formulation} we
briefly present the model and  formulation. The results and
discussions are presented in Sec. \ref{results}, and finally Sec.
\ref{conclusion} contains our conclusions.

\section{Model and Formulation}\label{formulation}

Thin film in the simple cubic geometry is treated in this work.
This film is layered structure which consist of interacting $L$ parallel layers.
Each of the layer has square lattice.  The Hamiltonian of the
isotropic Heisenberg model is given by
\eq{denk1}{\mathcal{H}=-\summ{<i,j>}{}{J_{ij}\paran{ s_i^xs_j^x+ s_i^ys_j^y+s_i^zs_j^z}}-\summ{i}{}{H_is_i^z}}
where $s_i^x,s_i^y$ and  $s_i^z$ denote the Pauli spin operators at a site $i$. $J_{ij}$ stands for the exchange interactions between the nearest
neighbor spins located at sites $i$ and $j$ and $H_i$ is the longitudinal magnetic field at a site $i$. The first sum is carried over the nearest neighbors of the thin film, while the second one is over all the sites. The exchange interaction $(J_{ij})$ between the spins on the sites $i$ and $j$ takes the values according to the positions of the nearest neighbor spins. Let we denote the intralayer exchange interactions in the surfaces of the film as $J_{1}$ and all other exchange interactions as $J_{2}$. This means that all nearest neighbor spins which belongs to the surfaces of the film interacted with $J_{1}$ with each other, while all other nearest neighbor spins have exchange interaction  $J_{2}$.

Magnetic fields are distributed according to a bimodal distribution
function to a lattice sites, which is given by:
\eq{denk2}{P\paran{H_i}=\frac{1}{2}\left[\delta\paran{H_i-H_0}+\delta\paran{H_i+H_0}\right]
} where $\delta$ stands for the delta function. This distribution distributes the to magnetic field $H_0$ half of the lattice sites and
$-H_0$ remaining half of the lattice sites randomly.

The simplest way for solving this system within the EFT formulation is using EFT-2 formulation \cite{ref42} which is two spin cluster approximation within the EFT formulation. This formulation is generalized form of the earlier formulation for the Ising model \cite{ref43}.
With following the same procedure given in Ref. \cite{ref23} we can arrive the magnetization expressions of each layer of the film as

% These expressions can be written  by using the differential operator technique and decoupling approximation (DA) \cite{ref45} as

\eq{denk3}{\begin{array}{lcl}
m_1&=&\sandd{\Theta_{1,1}^3\Theta_{2,2}}F_1\paran{x,y,H_0}|_{x=0,y=0}\\
m_k&=&\sandd{\Theta_{2,k-1}\Theta_{2,k}^3\Theta_{2,k+1}}F_2\paran{x,y,H_0}|_{x=0,y=0},k=2,3,\ldots,L-1\\
m_L&=&\sandd{\Theta_{2,L-1}\Theta_{1,L}^3}F_1\paran{x,y,H_0}|_{x=0,y=0}.\\
\end{array}} Here $m_i,(i=1,2,\ldots, L)$ denotes the magnetization of the $i^{th}$ layer. The operators in Eq. \re{denk3} are defined via
\eq{denk4}{
\Theta_{k,l}=\left[A_{kx}+m_lB_{kx}\right]\left[A_{ky}+m_lB_{ky}\right]
} where
\eq{denk5}{\begin{array}{lcl}
A_{km}&=&\cosh{\paran{J_k\nabla_m}}\\
B_{km}&=&\sinh{\paran{J_k\nabla_m}},\quad k=1,2; m=x,y.
\end{array}
}
The functions in Eq. \re{denk3} are given by
\eq{denk6}{F_n\paran{x,y,H_0}=\integ{}{}{}dH_1dH_2P\paran{H_1}P\paran{H_2}f_n\paran{x,y,H_1,H_2}} where
\eq{denk7}{f_n\paran{x,y,H_1,H_2}=\frac{\sinh{\paran{\beta X_0}}}{\cosh{\paran{\beta X_0}}+\exp{\paran{-2\beta J_n}}\cosh{\paran{\beta Y_0^{(n)}}}}} and where
\eq{denk8}{\begin{array}{lcl}
X_0&=&x+y+H_1+H_2\\
Y_0^{(n)}&=&\left[4J_n^2+(x-y+H_1-H_2)^2\right]^{1/2}\\
\end{array}}with the values $n=1,2$. In Eq. \re{denk7}, $\beta=1/(k_B T)$ where $k_B$ is Boltzmann
constant and $T$ is the temperature.

Magnetization expressions given in closed form in Eq. \re{denk3} can be constructed via acting differential operators on related functions. The effect of the exponential
differential operator to an arbitrary  function $G(x)$ is given by
\eq{denk9}{\exp{\paran{a\nabla}}G\paran{x}=G\paran{x+a}} with any
constant  $a$.
%%%%%%%%%%%%%%%%%%%%%%5

With the help of the Binomial expansion, Eq. \re{denk3} can be written in the form
\eq{denk10}{\begin{array}{lcl}
m_1&=&\summ{p=0}{6}{}\summ{q=0}{2}{}K_1\paran{p,q}m_1^{p} m_2^{q}\\
m_k&=&\summ{p=0}{2}{}\summ{q=0}{6}{}\summ{r=0}{2}{}K_2\paran{p,q,r}m_{k-1}^{p} m_k^{q}m_{k+1}^{r}\\
m_L&=&\summ{p=0}{6}{}\summ{q=0}{2}{}K_1\paran{p,q}m_L^{p} m_{L-1}^{q}\\
\end{array}}

where

\eq{denk11}{\begin{array}{lcl}
K_1(p,q)&=&\summ{i=0}{3}{}\summ{j=0}{3}{}\summ{k=0}{1}{}\summ{l=0}{1}{}k_1\paran{i,j,k,l}\delta_{p,i+j}\delta_{q,k+l}\\
K_2(p,q,r)&=&\summ{i=0}{1}{}\summ{j=0}{1}{}\summ{k=0}{3}{}\summ{l=0}{3}{}\summ{m=0}{1}{}\summ{n=0}{1}{}k_2\paran{i,j,k,l,m,n}\delta_{p,i+j}\delta_{q,k+l}\delta_{r,m+n}\\
\end{array}}
and

\eq{denk12}{\begin{array}{lcl}
k_1\paran{p,q,r,s}&=&\komb{3}{p}\komb{3}{q}A_{1x}^{3-p}A_{1y}^{3-q}A_{2x}^{1-r}A_{2y}^{1-s}B_{1x}^{p}B_{1y}^{q}B_{2x}^{r}B_{2y}^{s}F_1\paran{x,y,H_1,H_2}|_{x=0,y=0}\\
k_2\paran{p,q,r,s,t,v}&=&\komb{3}{r}\komb{3}{s}A_{2x}^{5-(p+r+t)}A_{2y}^{4-(q+s+v)}B_{2x}^{p+r+t}B_{2y}^{q+s+v}F_2\paran{x,y,H_1,H_2}|_{x=0,y=0}.\\
\end{array}}
These coefficients can be calculated from the definitions given in Eq. \re{denk5} with using Eqs. \re{denk6} and \re{denk9}.

For a given Hamiltonian parameters and temperature, by determining the coefficients  from Eq. \re{denk11} we can obtain a system of coupled non linear equations from Eq. \re{denk10}, and by solving this system we can get the longitudinal magnetizations of each layer ($m_i,i=1,2,\ldots,L$). The solution of that equation system can be done in a numerical way e.g. with using usual Newton-Raphson method.  The total longitudinal magnetization ($m$) can be calculated via
\eq{denk13}{m=\frac{1}{L}\summ{i=1}{L}{m_i}.}

Since all longitudinal magnetizations are close to zero in the vicinity of the second order critical point, we can obtain another coupled  equation system to determine the transition temperature by linearizing the equation system given in  Eq. \re{denk10}, i.e.
\eq{denk14}{\begin{array}{lcl}
m_1&=&K_1\paran{1,0}m_1+K_1\paran{0,1}m_2\\
m_k&=&K_2\paran{1,0}m_{k-1}+K_2\paran{0,1,0}m_{k}+K_2\paran{0,0,1}m_{k+1}\\
m_L&=&K_1\paran{1,0}m_L+K_1\paran{0,1}m_{L-1}.\\
\end{array}}

Critical temperature ($T_c$) can be determined from $\mathbf{\mathrm{det(A)=0}}$ where $A$ is the matrix of coefficients of the linear equation system given in Eq. \re{denk14}.

\section{Results and Discussion}\label{results}

Let us choose unit of energy as $J_2>0$ and scale the temperature ($k_BT$) and magnetic field ($H_0$) as well as $J_1$ with $J_2$,
\eq{denk15}{
r=\frac{J_1}{J_2}, t=\frac{k_BT}{J_2}, h_0=\frac{H_0}{J_2},
} Since we are interested in only the ferromagnetic case, the parameter $r$  is positive or zero.

First, let us investigate the variation of the critical temperature of the thin film with the magnetic field $h_0$, in the case of $r=1$. Since the field distribution $\pm h_0$ present, then it is expected that the critical temperature of the film decreases with rising $h_0$. Rising $h_0$ drags the system to the disordered state and when this effect (for a given temperature) can overcome to the spin-spin interaction (which trying to keep the system in an ordered phase), system passes to the disordered phase.  This behavior can be seen in Fig. \re{sek1} for several values of the film thickness. As seen in the Fig. \re{sek1}, rising $h_0$ causes the decline of the critical temperature. Besides, for a fixed value of $h_0$ thicker films have higher critical temperature than the thinner ones. Phase diagrams terminates at a certain $h_0$ value. There should be a tricritical behavior, since the phase diagrams terminates at a finite values of critical temperature. The coordinate of the tricritical point is $\paran{t_c,h_0}=\paran{2.000,2.670}$ for $L=6$. The same value for the corresponding bulk system (simple cubic lattice) is  $\paran{t_c,h_0}=\paran{2.274,2.748}$ \cite{ref41}. As seen in Fig. \re{sek1}, both of the $h_0$ and $t_c$ coordinates of the tricritical point rises when the film thickness rises. This rising trend for the $\paran{t_c,h_0}$ coordinates of the tricritical point with rising film thickness can also be verified by comparing them with the coordinates of the bulk system. When the film gets thicker, the physical properties of the film 	approaches to the bulk system.

\begin{figure}[h]\begin{center}
\epsfig{file=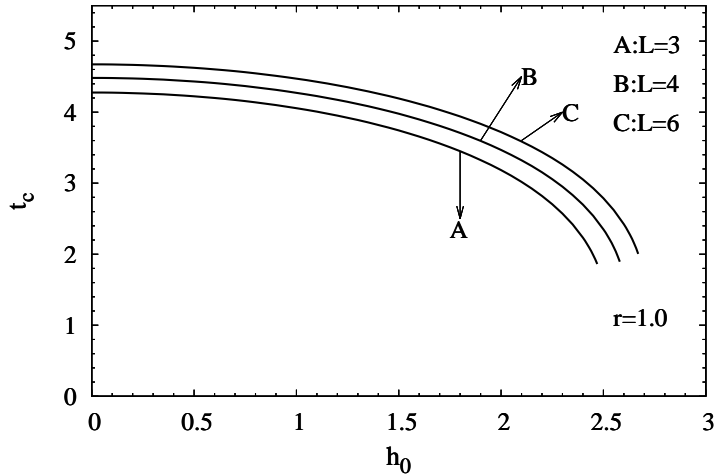, width=10cm}
\end{center}
\caption{Variation of the critical temperature with the  magnetic field (center of the distribution) in the $(t_c-h_0)$ plane for the isotropic Heisenberg  model in the thin film geometry, for some selected values of the film thickness.
} \label{sek1}\end{figure}

In order to see this tricritical behavior, let us look at the variation of the  magnetization with temperature at this higher values of $h_0$. The variation of the magnetization with the temperature can be seen in Fig. \re{sek2}, for the film thickness $L=6$. In each figure, total magnetization $m$, surface magnetization $m_1$  and the magnetization of the inner layer $m_3$ present. As we can see from the Fig. \re{sek2} that, for all the values of the $h_0$, surface magnetization is lower than the magnetization of the inner layer. This fact comes from the excess in the number of nearest neighbor of the spins located at the inner layers. Besides, we can see from the Fig. \re{sek2} (d) that, the transition from the ferromagnetic phase to the paramagnetic phase is discontinuous. In other words, for the higher values of the $h_0$, first order transitions occur. This fact is consistent with the phase diagrams present in Fig. \re{sek1}. If we look at the Figs. \re{sek2} (a)-(d) we can see that, when the $h_0$ rises, phase transitions transform from the second order type to the first order type.
Lastly, as we see from the Fig. \re{sek2} (d), the ground state of the surface layer is not completely ordered in contrast to the inner layer (compare the curves related $m_1$ and $m_3$). Magnetic field distribution destroys the completely ordered ground state of the surface layer. Then due to the thermal
agitations which occurs with rising temperature, magnetization of the surface layer rises for a while then it starts to decrease. This induce the cusp like behavior of the total magnetization as seen in Fig. \re{sek2} (d). Besides, we can say that, rising $h_0$ first destroys the order of the surface and after then the inner layers.

\begin{figure}[h]\begin{center}
\epsfig{file=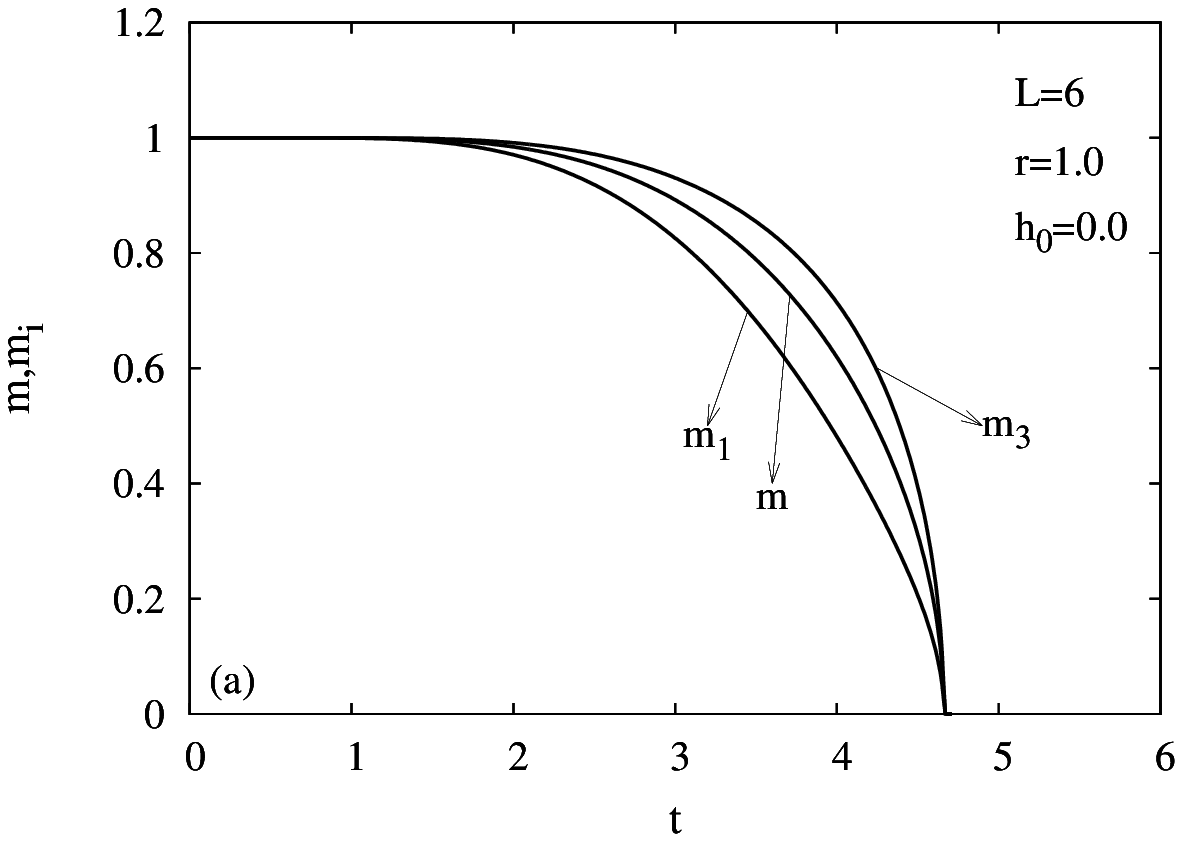, width=6cm}
\epsfig{file=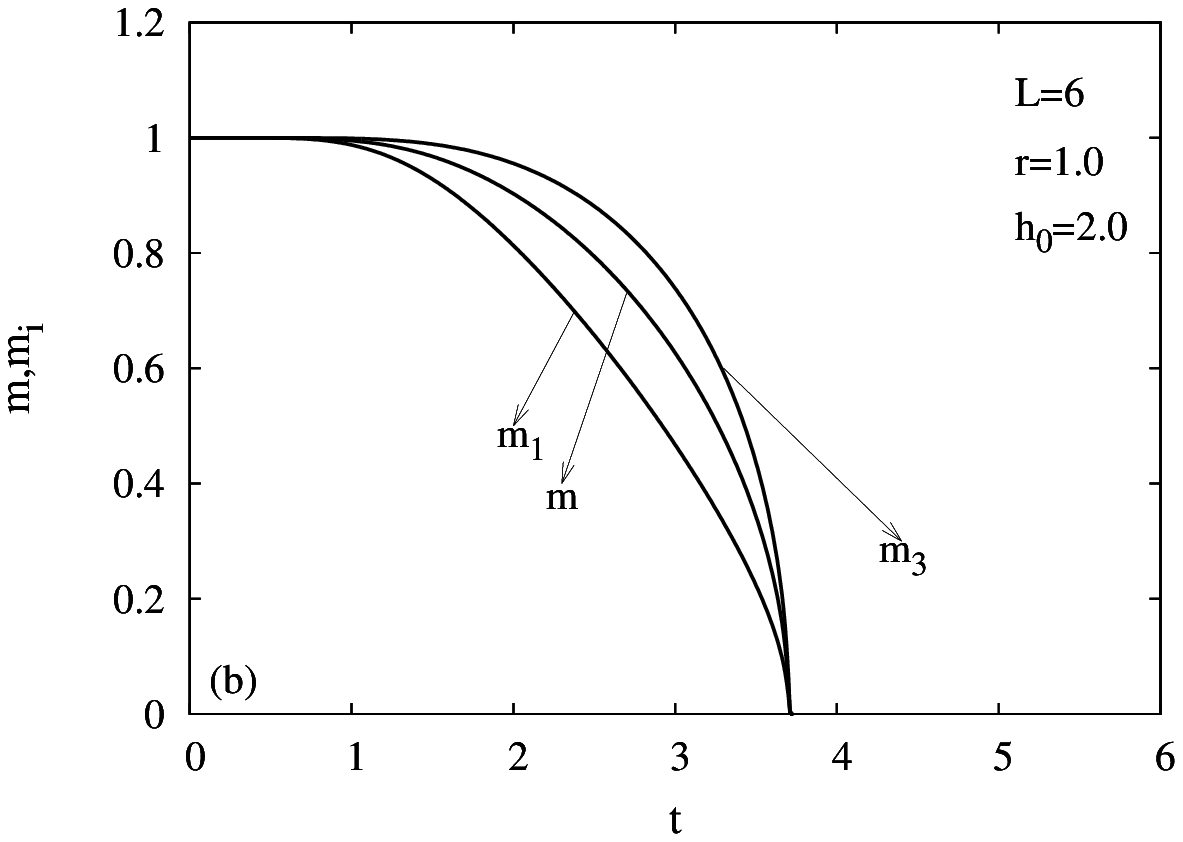, width=6cm}
\epsfig{file=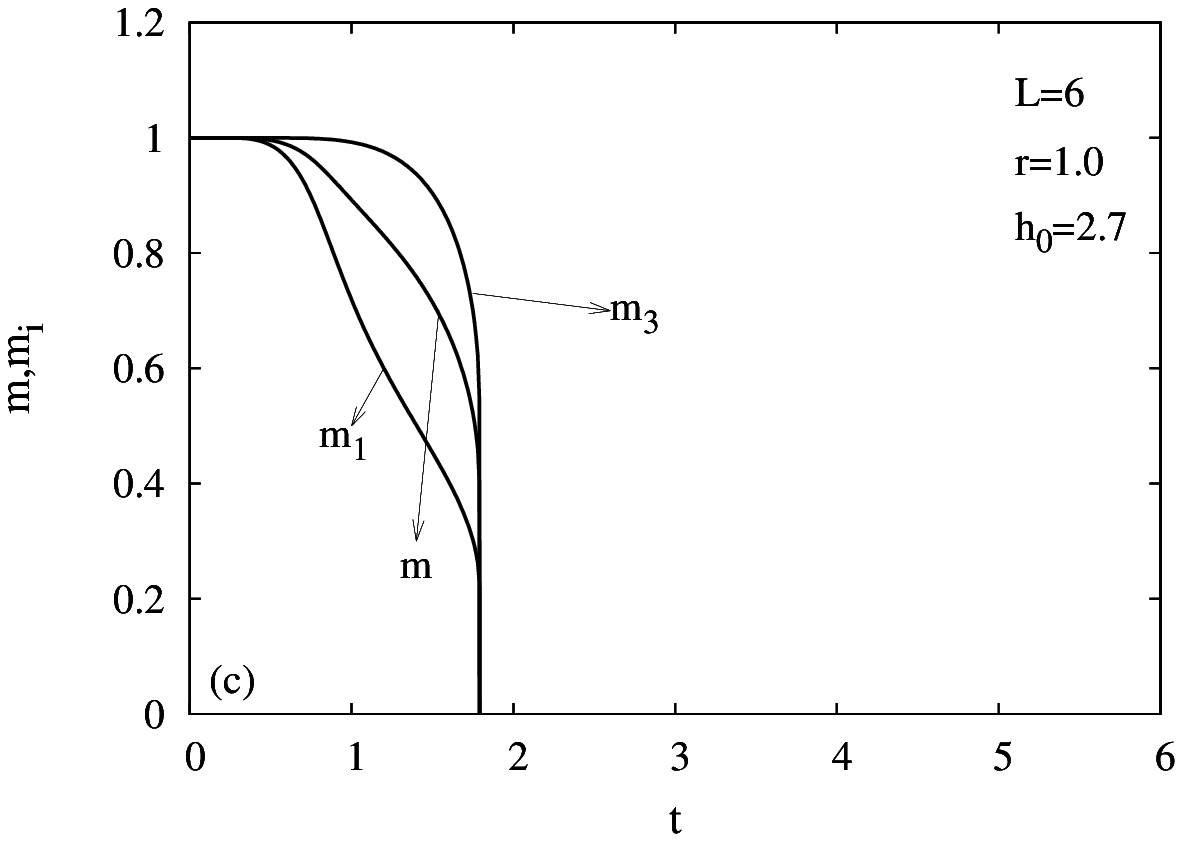, width=6cm}
\epsfig{file=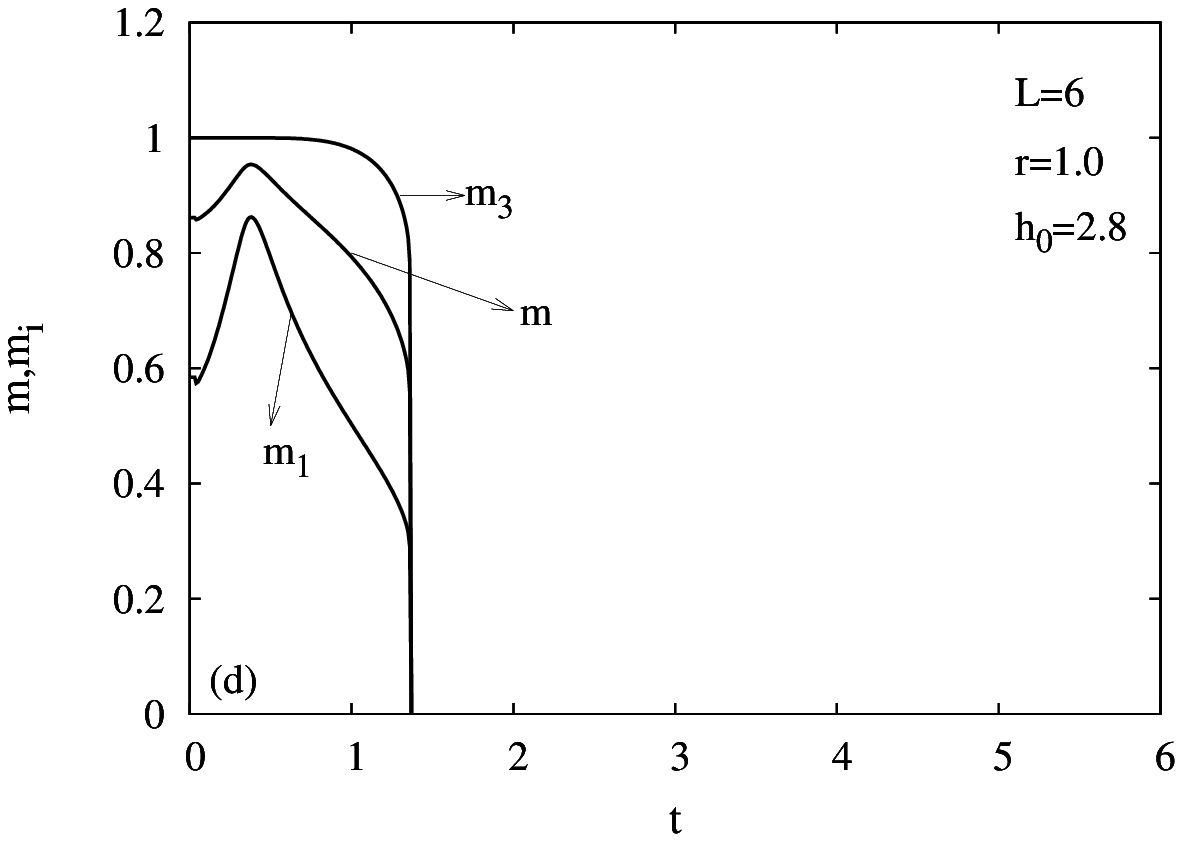, width=6cm}
\end{center}
\caption{Variation of the magnetization of the surface layer ($m_1$), inner layer ($m_3$), as well as the total magnetization ($m$) of the thin film with the temperature, for some selected values of $h_0$, with film thickness $L=6$.} \label{sek2}\end{figure}

Now, let us look what is happening when the surface layers have greater value of exchange interaction than the inner layers, i.e. the case $r>1$. Variation of the critical temperature with $r$ can be seen in Fig. \re{sek3} for different values of $L$ and $h_0$.
It is well established both theoretically and experimentally fact that,  for the systems with a surface, magnetically disordered surface can coexist with a magnetically ordered bulk phase for the values of $r$ that provide $r<r^{*}$ while the reverse can occur for the values $r>r^{*}$, i.e. surface can reach the magnetically ordered phase before the bulk. This $r^{*}$ point is called special point. For the values of $r<r^{*}$ thicker films have higher critical temperature than the thinner ones. Otherwise, the reverse relation holds. This fact can be seen in Fig. \re{sek3}.  For the special point coordinate in the absence of the magnetic field ($h_0=0.0$)  we find $(t_c^{*},r^{*})=(4.8910,1.3454)$, where the first one is just the
critical temperature of the corresponding bulk system (the system with simple cubic lattice) in a same model \cite{ref42}. The value of $1.3454$ can be compared with the Ising counterpart which is obtained as $1.3068$ with EFT with differential operator technique\cite{ref44}.

\begin{figure}[h]\begin{center}
\epsfig{file=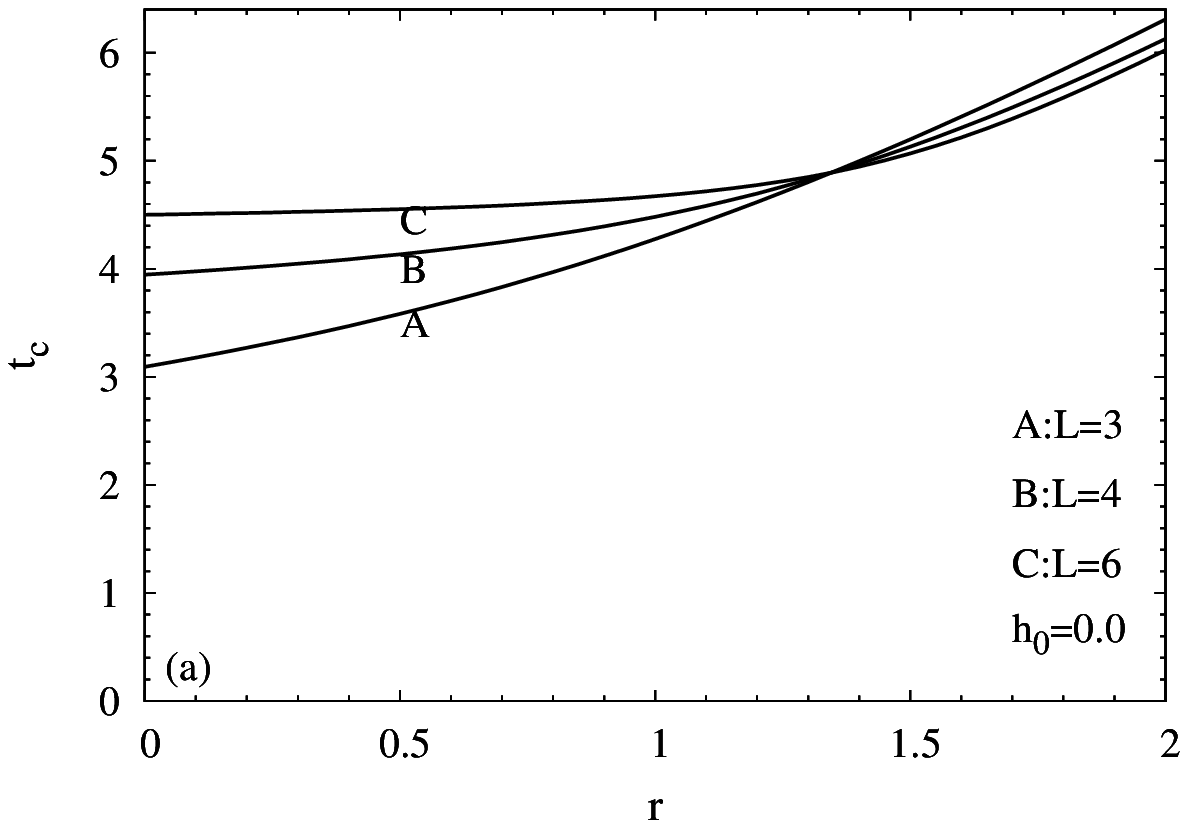, width=6cm}
\epsfig{file=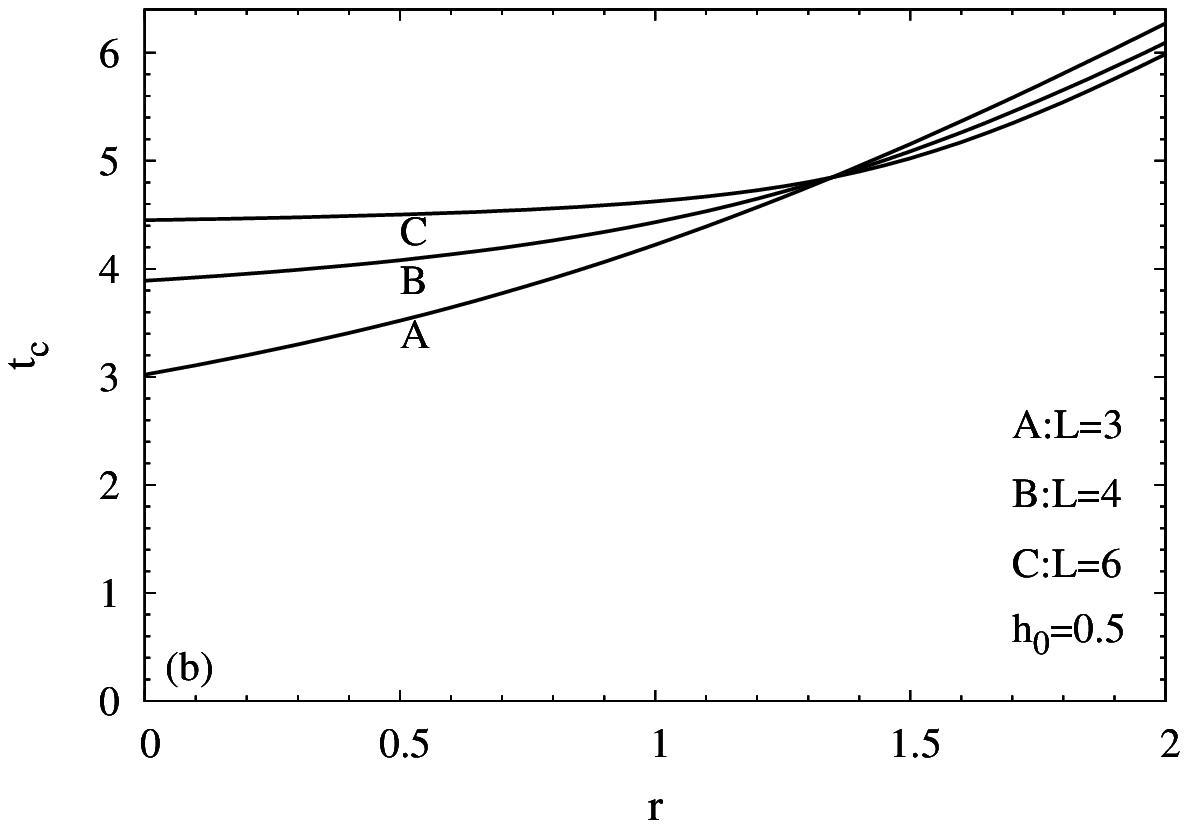, width=6cm}
\epsfig{file=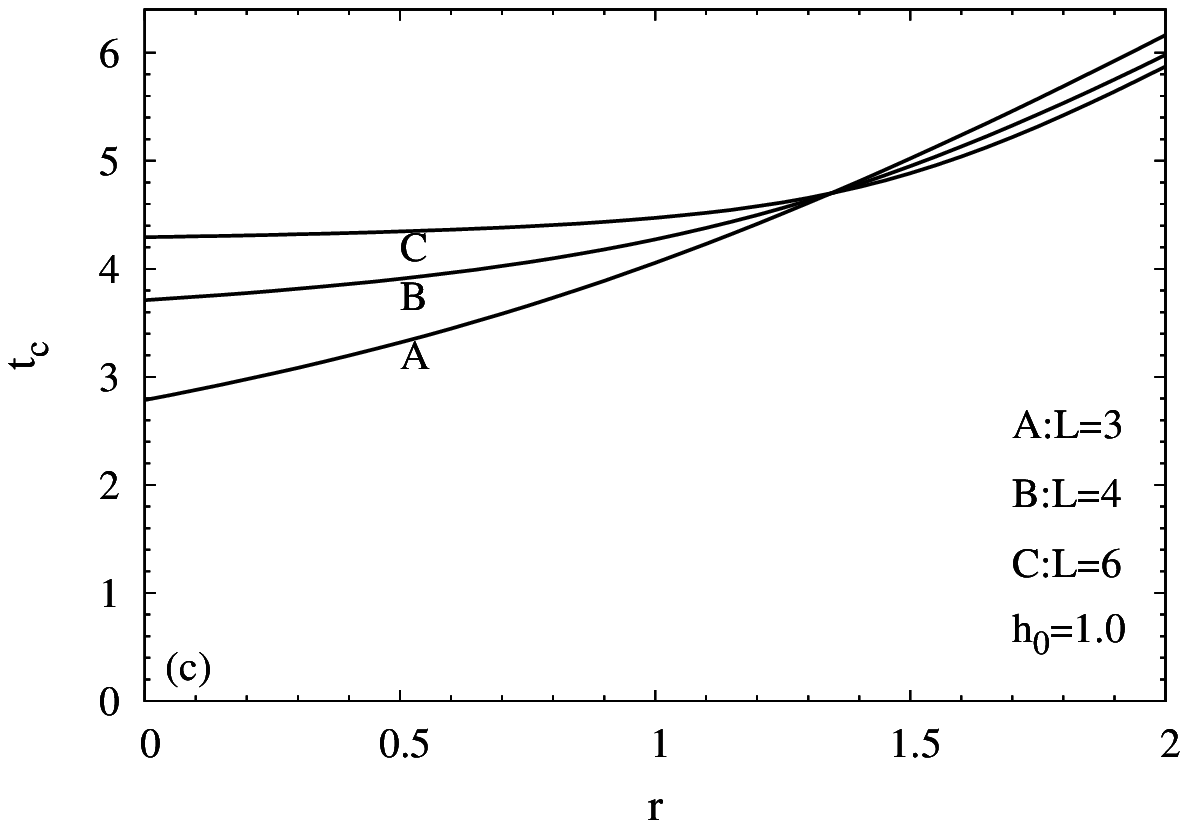, width=6cm}
\epsfig{file=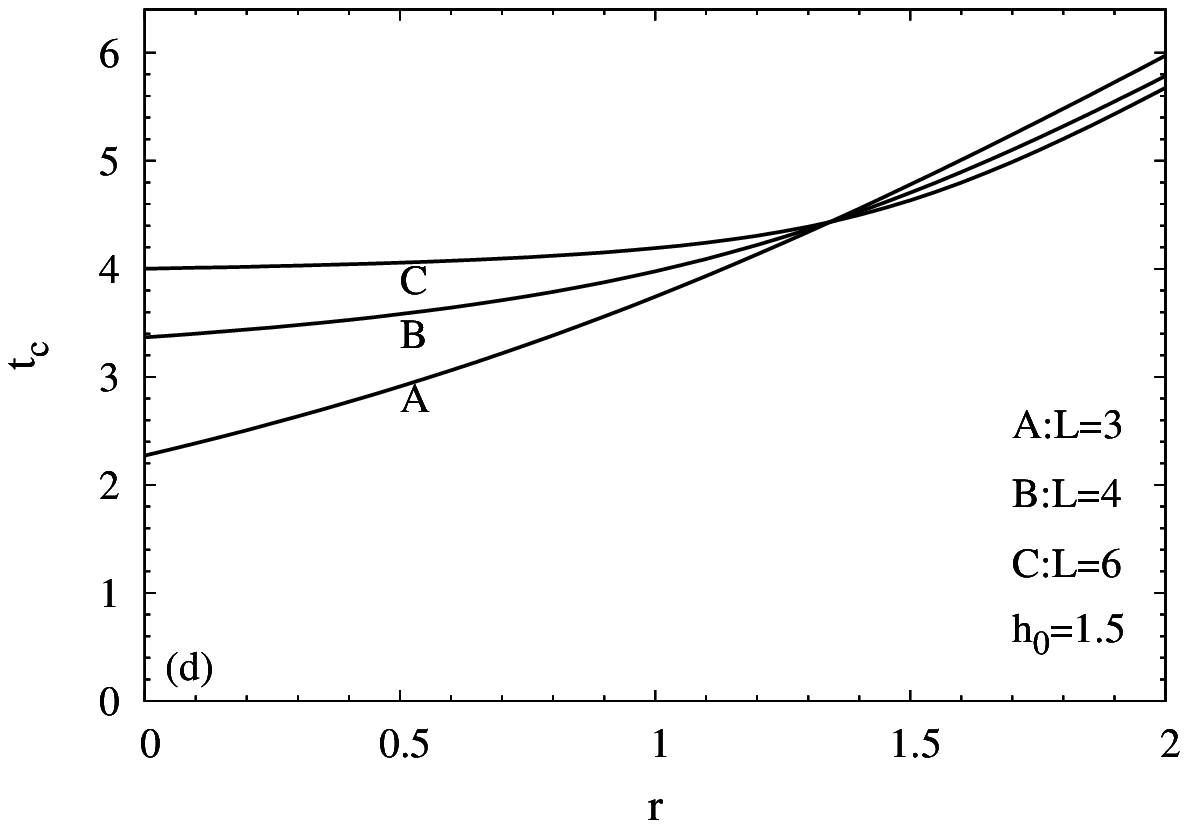, width=6cm}
\epsfig{file=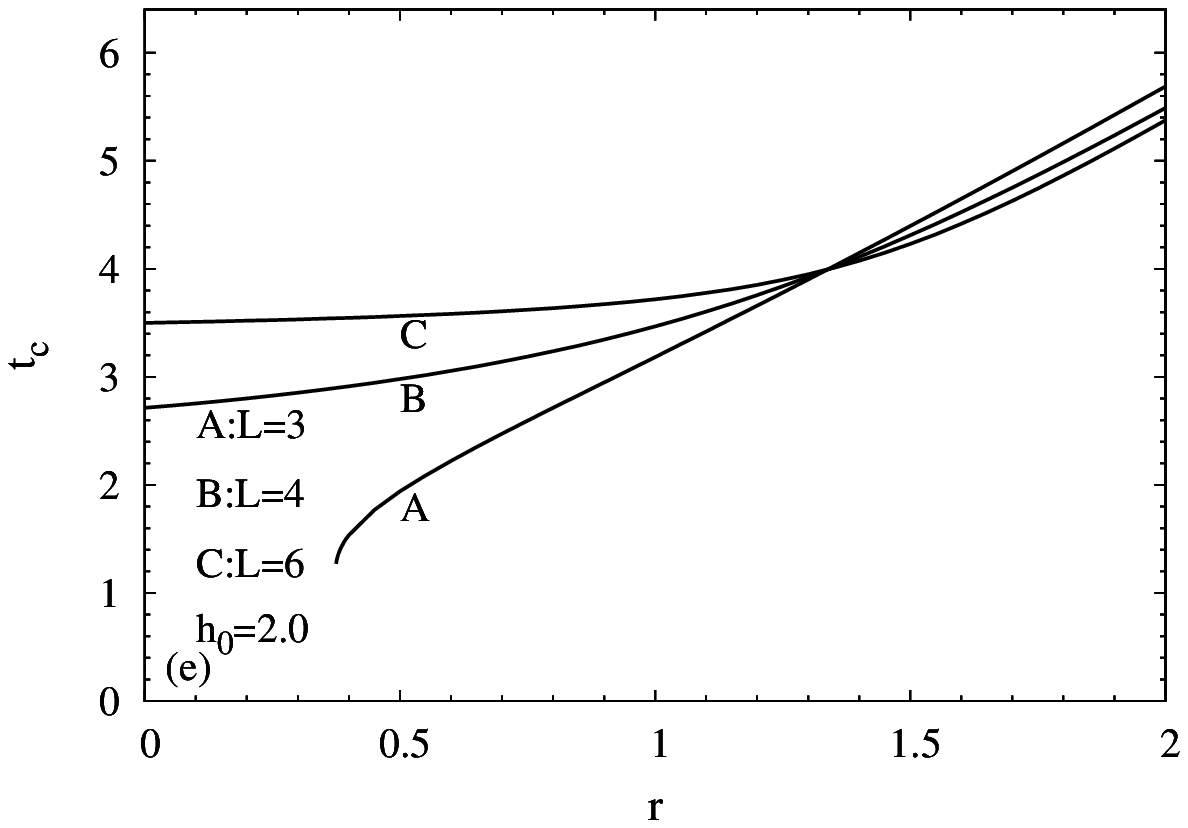, width=6cm}
\epsfig{file=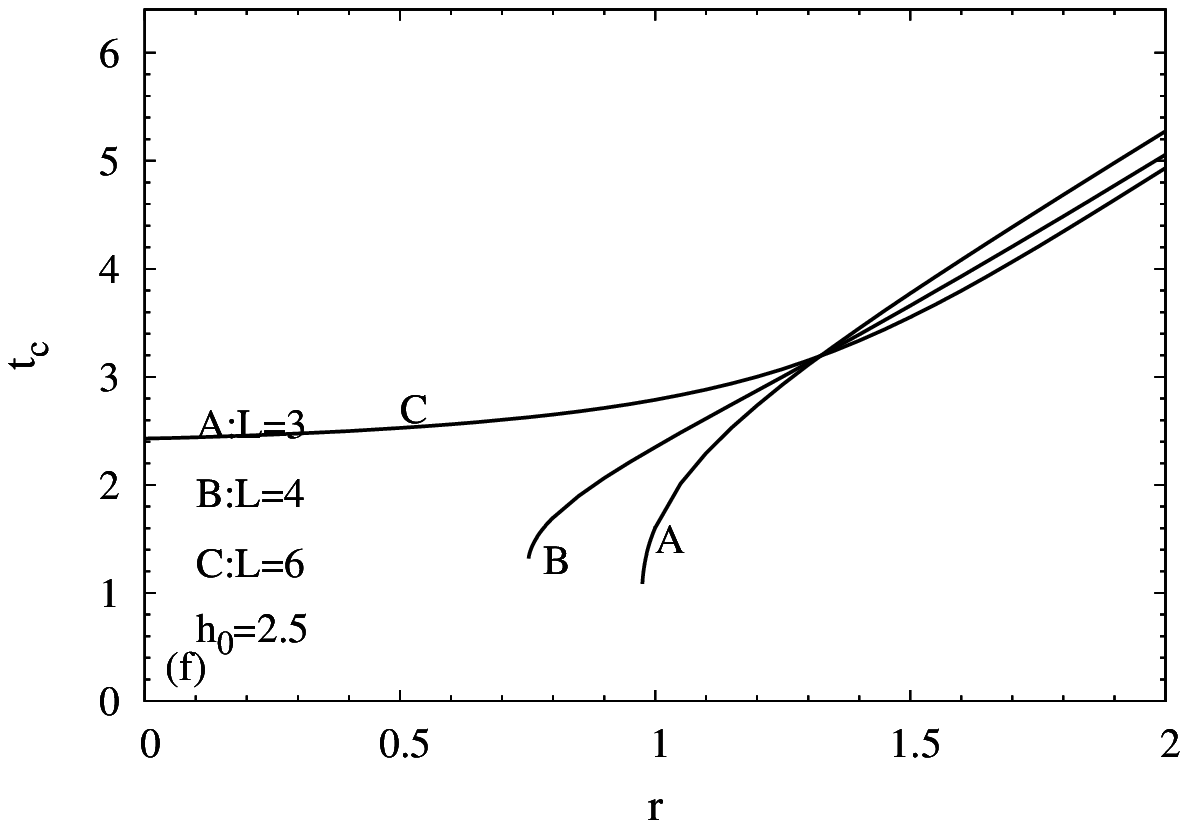, width=6cm}
\epsfig{file=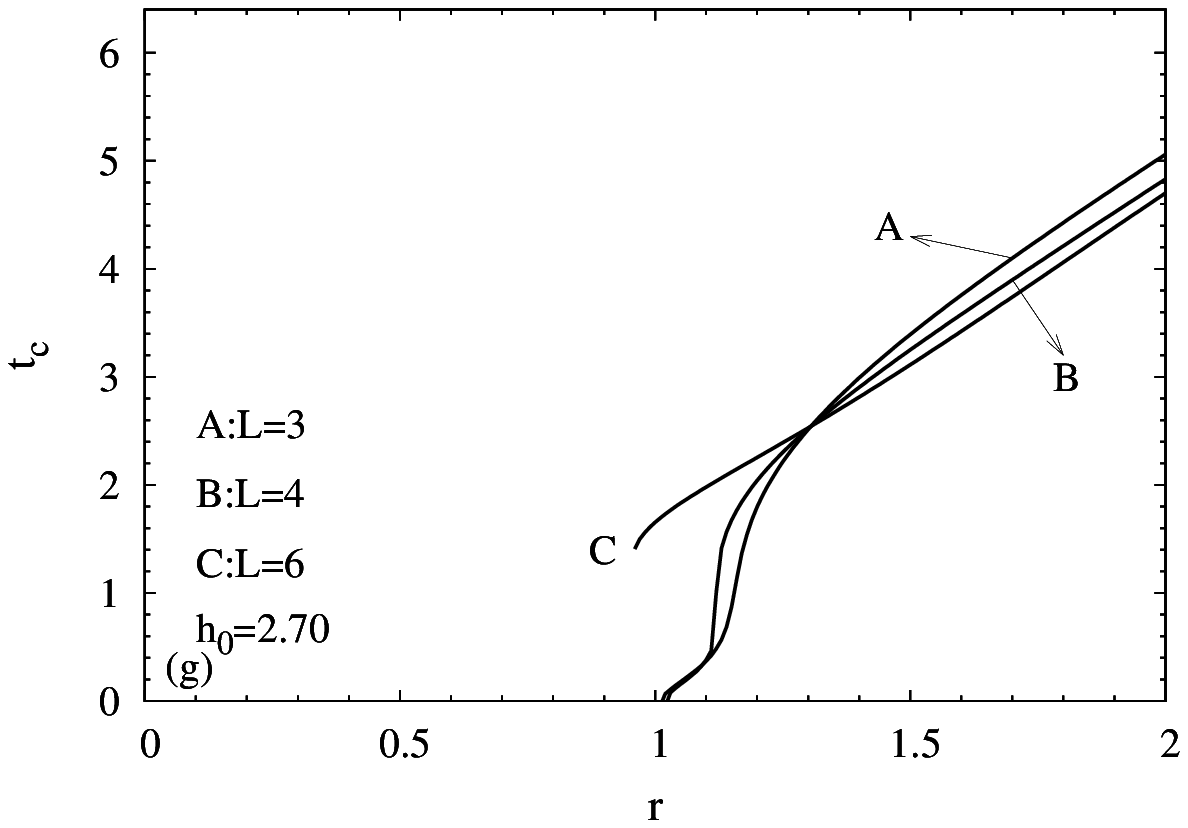, width=6cm}
\epsfig{file=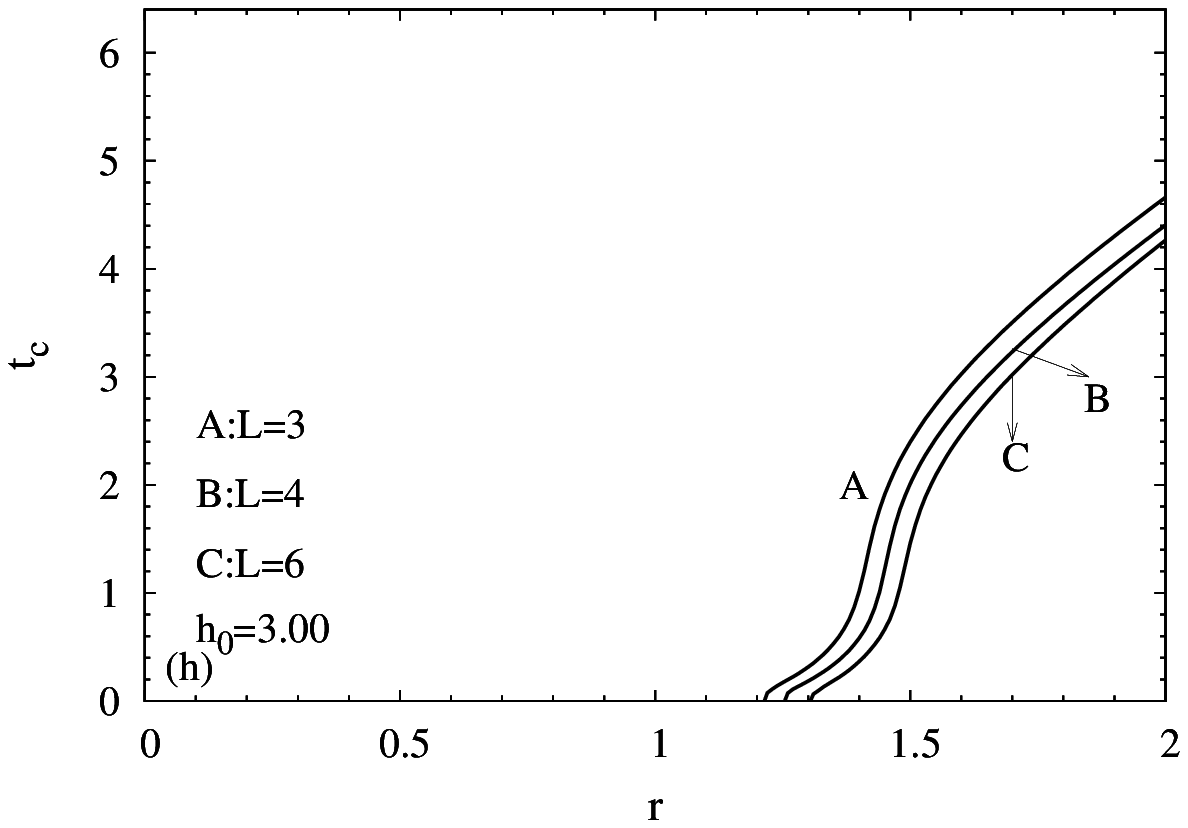, width=6cm}
\end{center}
\caption{Variation of the critical temperature with  $r$ for some selected values of film thickness and magnetic field.} \label{sek3}\end{figure}

When $h_0$ rises, the special point coordinate $r^{*}$ does not  change significantly, while the corresponding value of the critical temperature $t_c^{*}$ decreases. $r^*$ coordinate of the special point when $h_0=0.0$ is $r^*=1.3454$ and value of the $r^*$ coordinate of the special point changes as $r^*=1.3036$, when right before it disappears. The variation of the  $t_c^{*}$ with $h_0$ can be seen in Fig. \re{sek4} (a). At the same time, this curve corresponds to the variation of the critical temperature of the simple cubic lattice with $h_0$. This curve terminates at a value $(t_c^{*},h_0)=(2.274,2.748)$. After the value of $h_0=2.748$, thin film can not have special point.  This point is nothing but the tricritical point of the isotropic quantum Heisenberg model on a simple cubic lattice with a bimodal random magnetic field distribution \cite{ref41}.  The absence of the special point can also be seen in Fig. \re{sek3} (h). The typical effect of the rising $h_0$ on the curves in the $(t_c-r)$ plane can be seen in e.g. curve labeled by C in Figs. \re{sek3} (e)-(h). Rising $h_0$ first shifts the whole curve downward (compare curves labeled by C in Figs. \re{sek3} (e) and (f)). After a specific value of $h_0$ (which depends on the film thickness) curves starts to terminate at a tricritical point on the right side (curve labeled by C in Fig. \re{sek3} (g)). At this point special point still exist, i.e. for $r<r^*$ thicker films have higher critical temperature and vice versa. If $h_0$  continue to increase, then special point cannot survive and the curves related to thicker film wholly settle to the under of the thinner film  (e.g. compare curves labeled by C and B in Fig. \re{sek3} (h)). At the same time tricritical point disappears during this last step.

Another interesting property namely tricritical point changes while $h_0$ rises. When  $h_0$ rises, tricritical point first appears in thinner films (compare curves labeled by A and B in Fig. \re{sek3} (e)). Then, after a specific value of the $h_0$ (which depends on the $L$) tricritical point disappears. During this process, no significant change in the $t_c$ coordinate of the tricritical point has been observed. On the other hand the variation of the $r$ coordinate of the tricritical point changes and this can be seen in Fig. \re{sek4} (b). As seen in \re{sek4} (b), the region that have tricritical behavior in the $(r-h_0)$ plane is wider for the thinner film than the thicker one  (compare curves labeled by A and B in Fig. \re{sek4} (b)).

\begin{figure}[h]\begin{center}
\epsfig{file=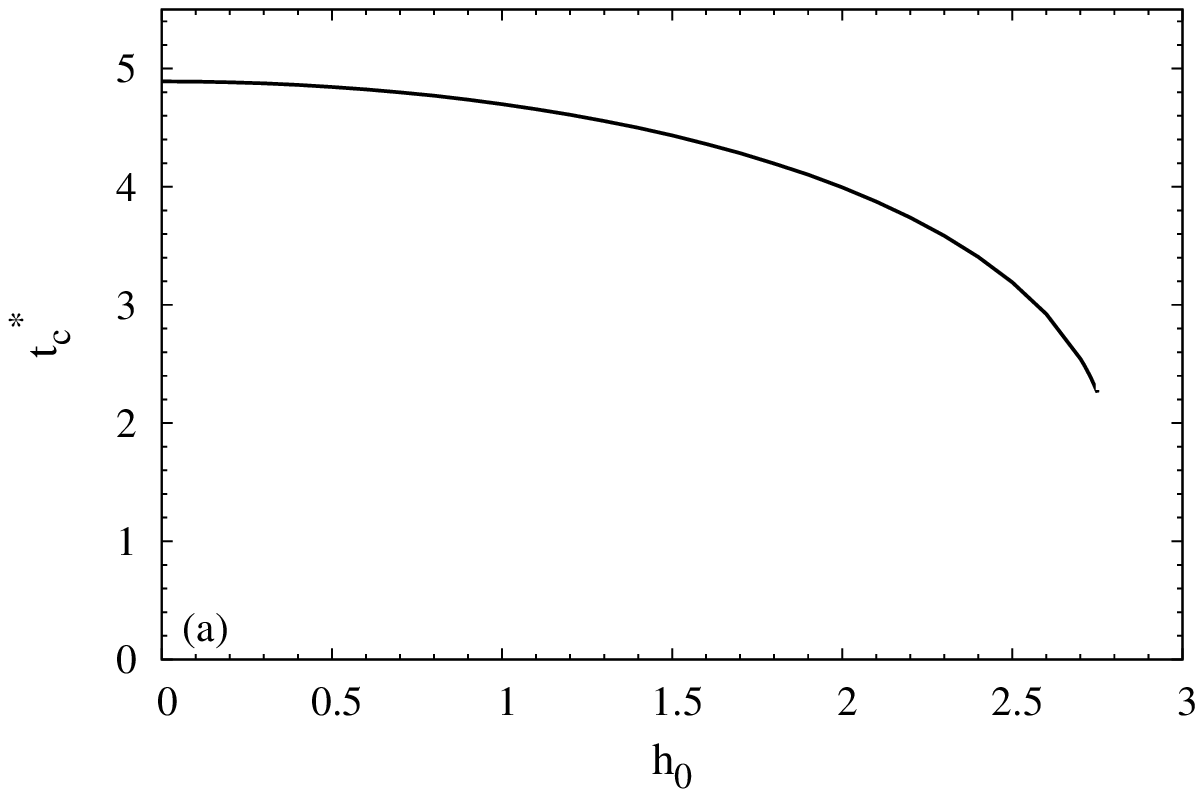, width=6cm}
\epsfig{file=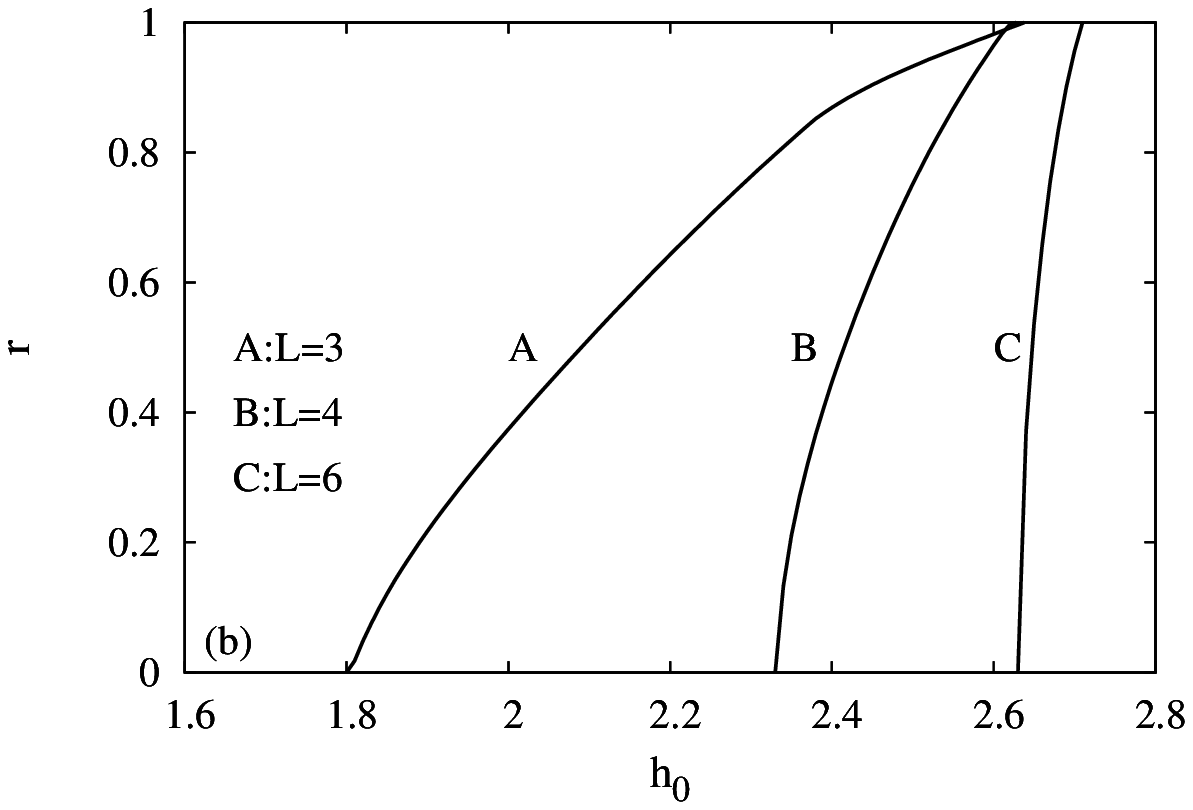, width=6cm}
\end{center}
\caption{(a)Variation of the temperature coordinate of the special point $(t_c^*)$ with the magnetic field ($h_0$), (b) Variation of the $r$ coordinate of the tricritical point  with the magnetic field ($h_0$) for different thickness values.
} \label{sek4}\end{figure}

The relation between the critical temperature of the film and the corresponding bulk system shows itself also in the
relation between the magnetization of the surface layers and inner layers. In other words, chosen $r$ also determines the relation between the magnetization of the surface layers and inner layers for any temperature, which is below the critical temperature. In order to more elaborate on this point, the variation of the magnetization of the surface and inner layer as well as the total magnetization of the film with the temperature can be seen in Figs. \re{sek5}-\re{sek8} for the film thickness $L=3,6$  and several values of $h_0$.
We choose $r=0.1<r^{*}$ in Figs. \re{sek5} and \re{sek7} and $r=2.2>r^{*}$ in Figs. \re{sek6} and \re{sek8}.

\begin{figure}[h]\begin{center}
\epsfig{file=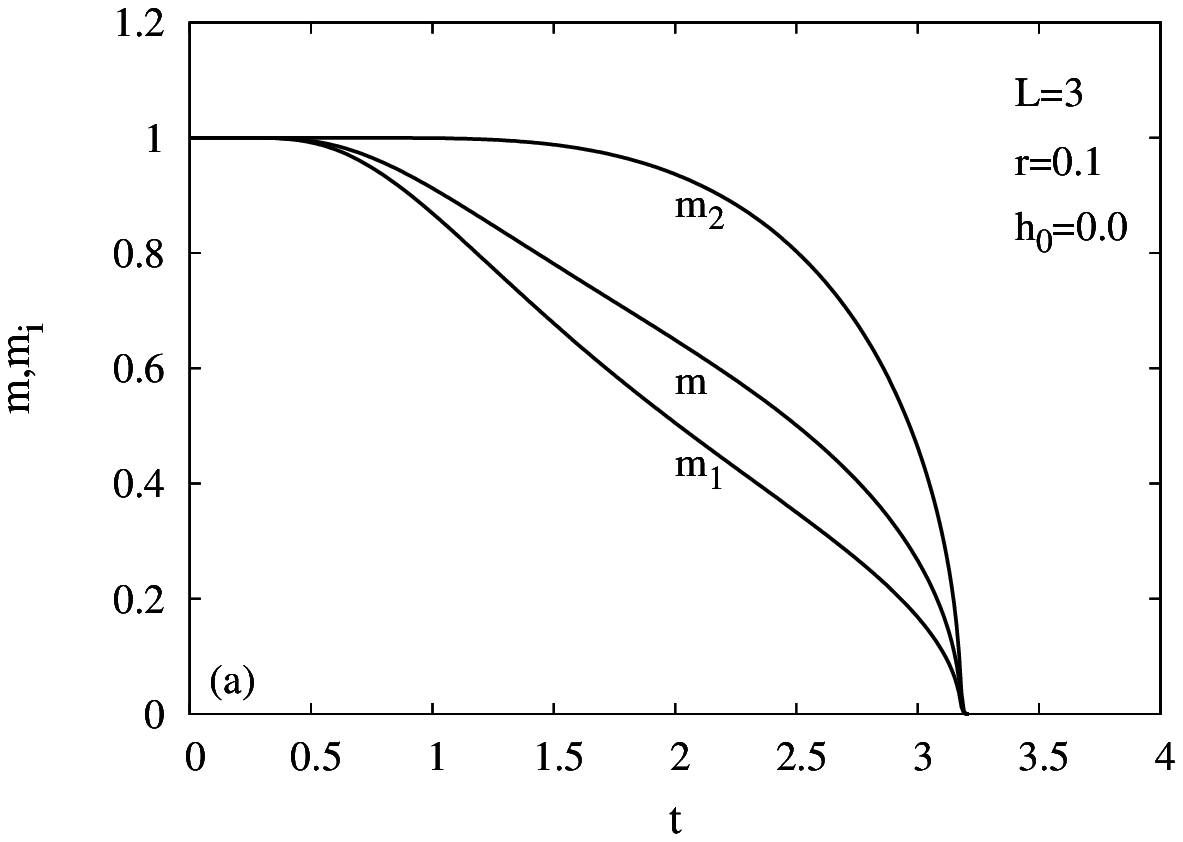, width=6cm}
\epsfig{file=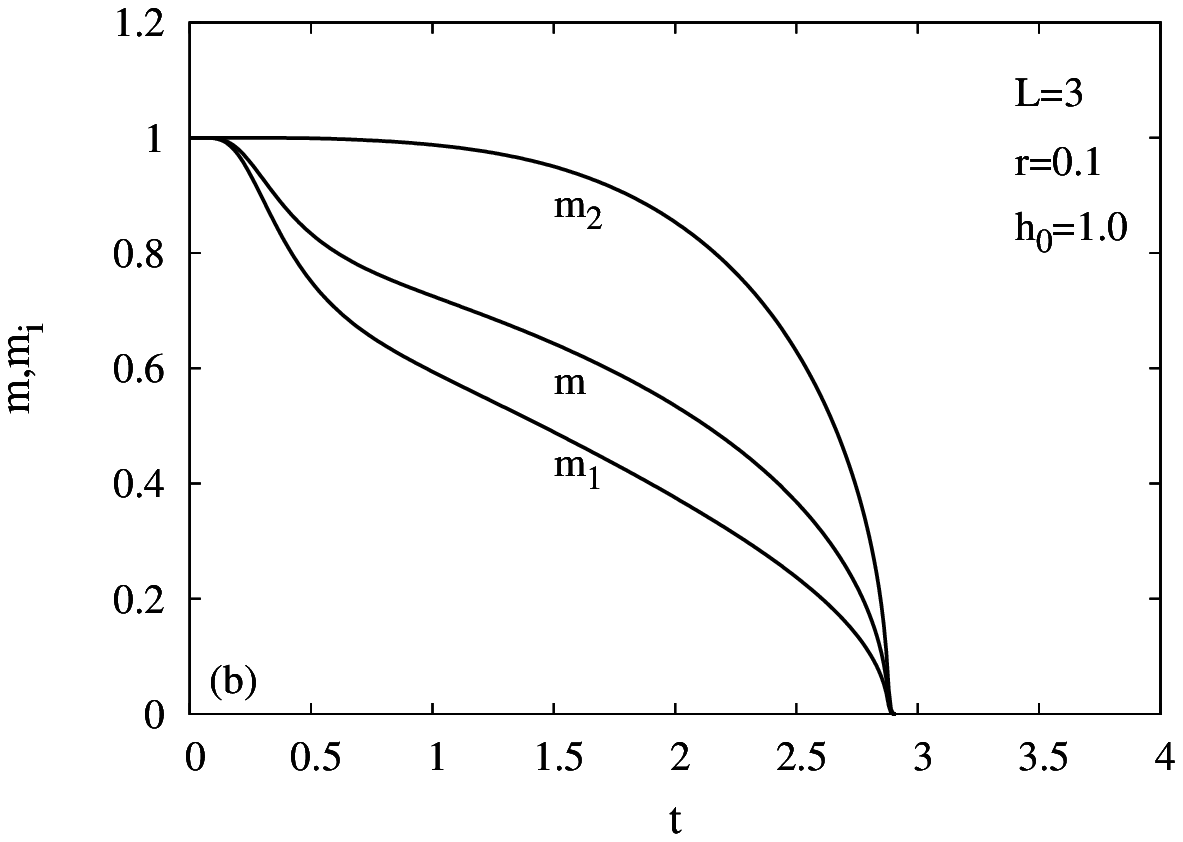, width=6cm}
\epsfig{file=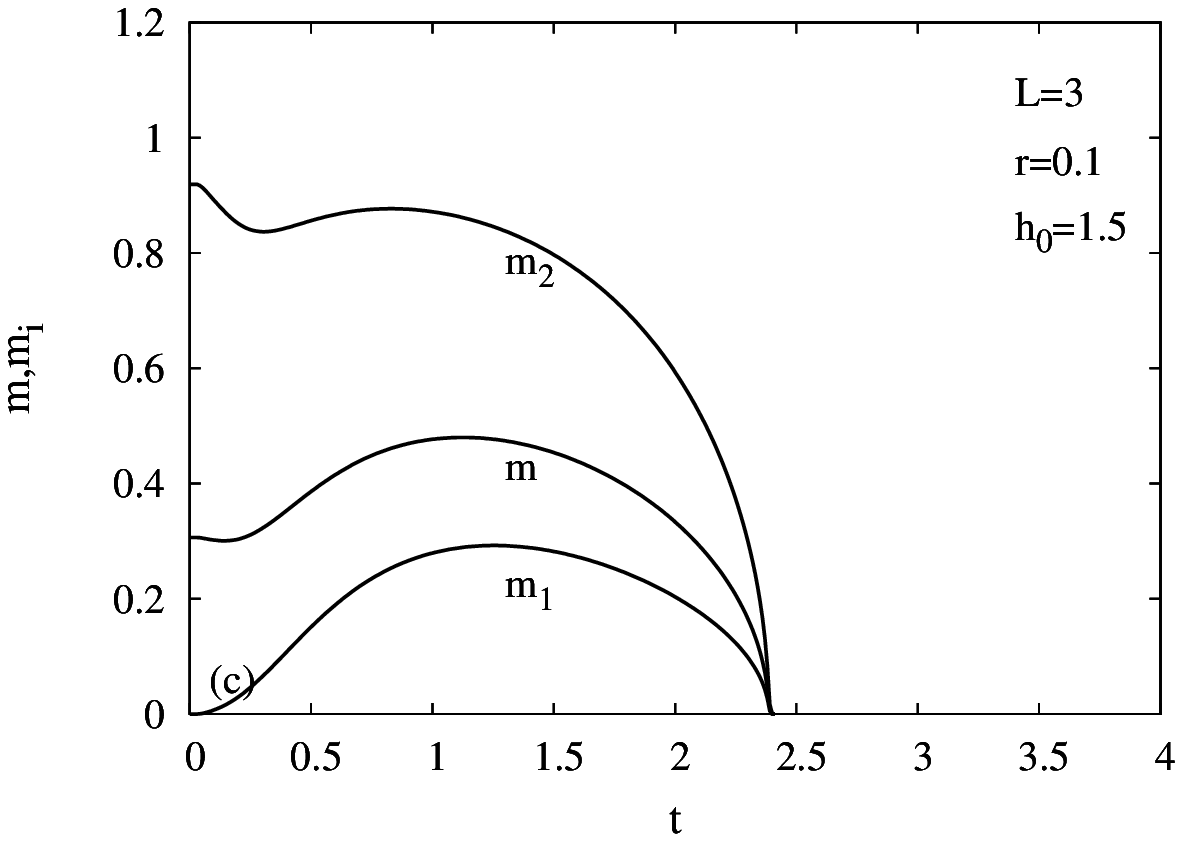, width=6cm}
\epsfig{file=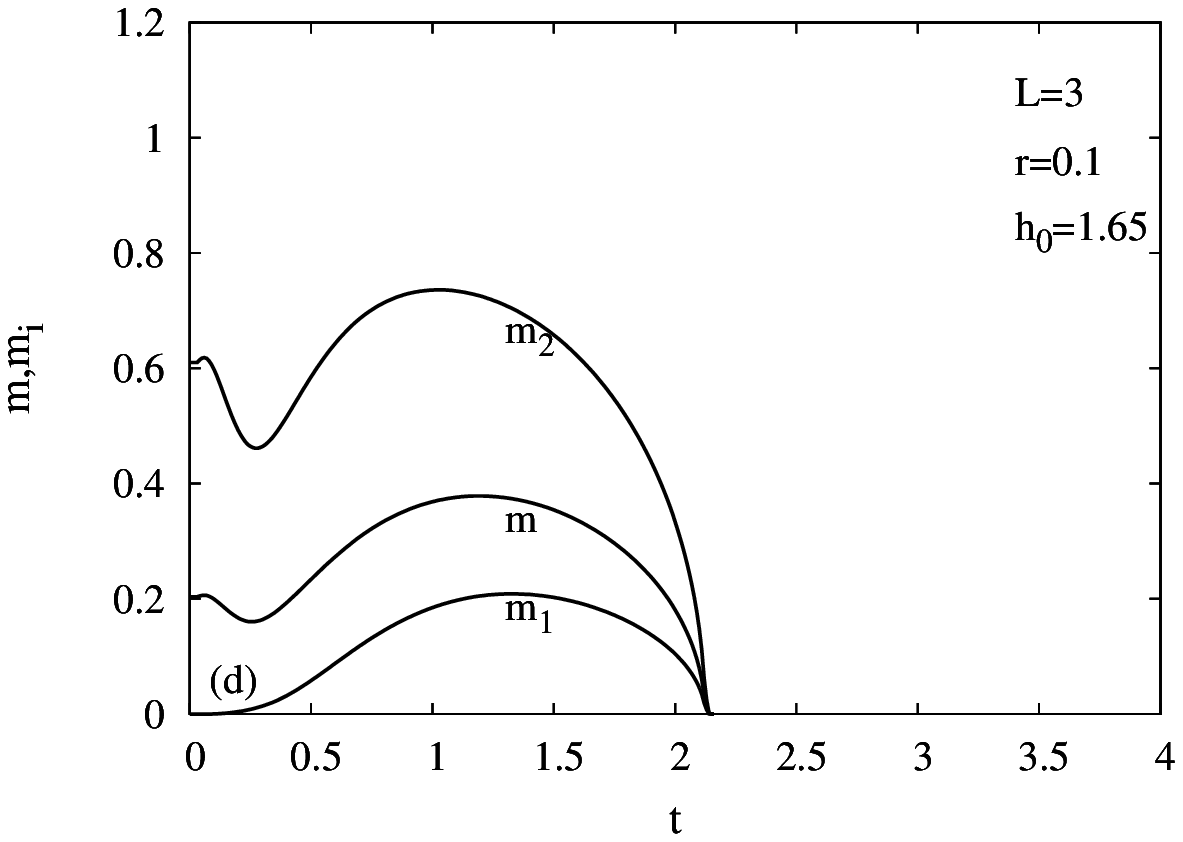, width=6cm}
\end{center}
\caption{Variation of the magnetization of the surface layer ($m_1$), inner layer ($m_2$), as well as the total magnetization ($m$) of the thin film with thickness $L=3$, with the temperature, for some selected values of $h_0$.  $r$ value is fixed as $r=0.1<r^*$. } \label{sek5}\end{figure}

\begin{figure}[h]\begin{center}
\epsfig{file=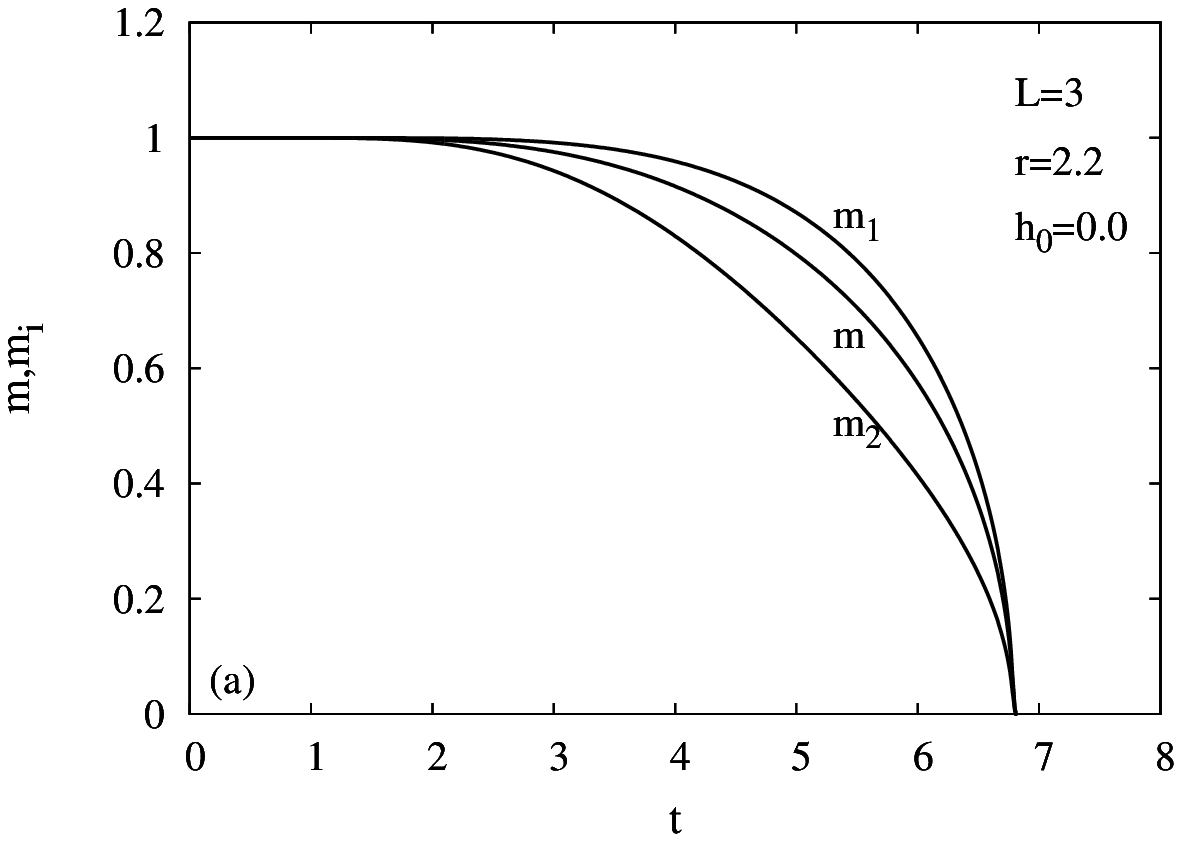, width=6cm}
\epsfig{file=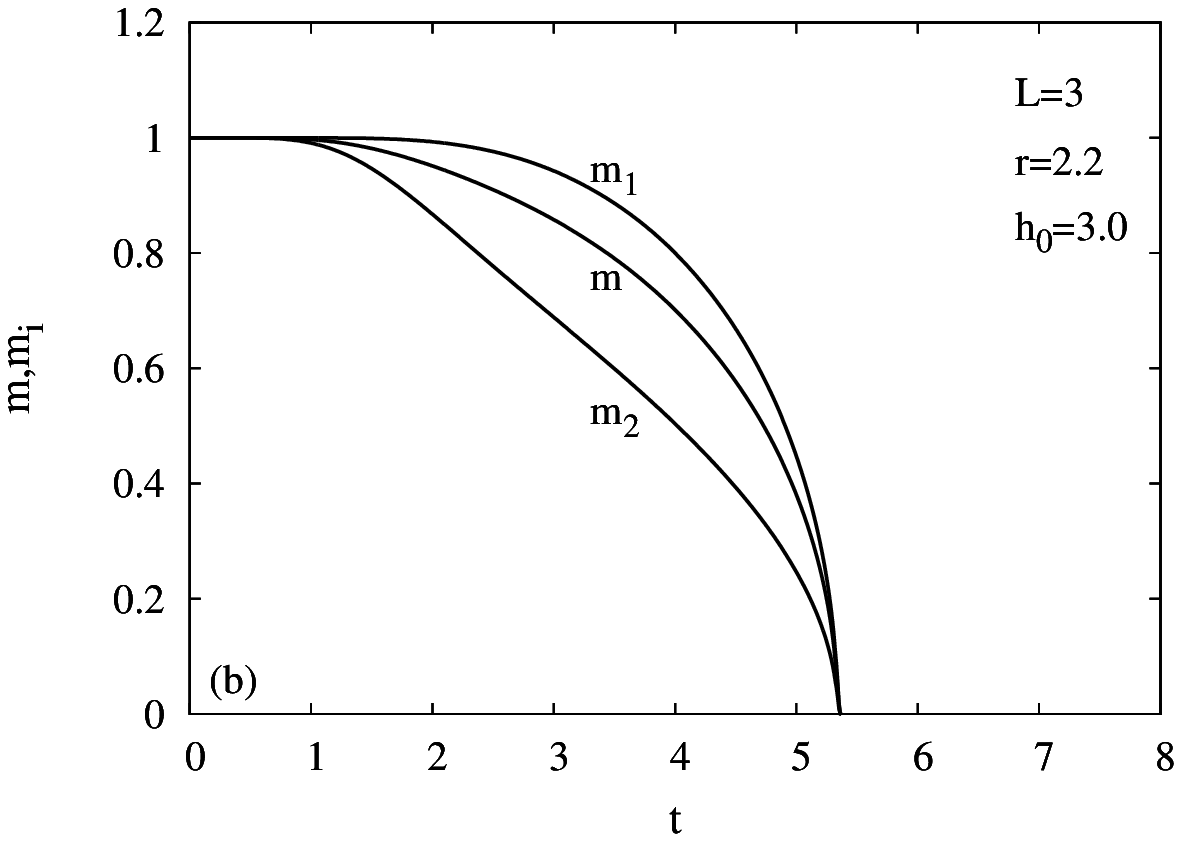, width=6cm}
\epsfig{file=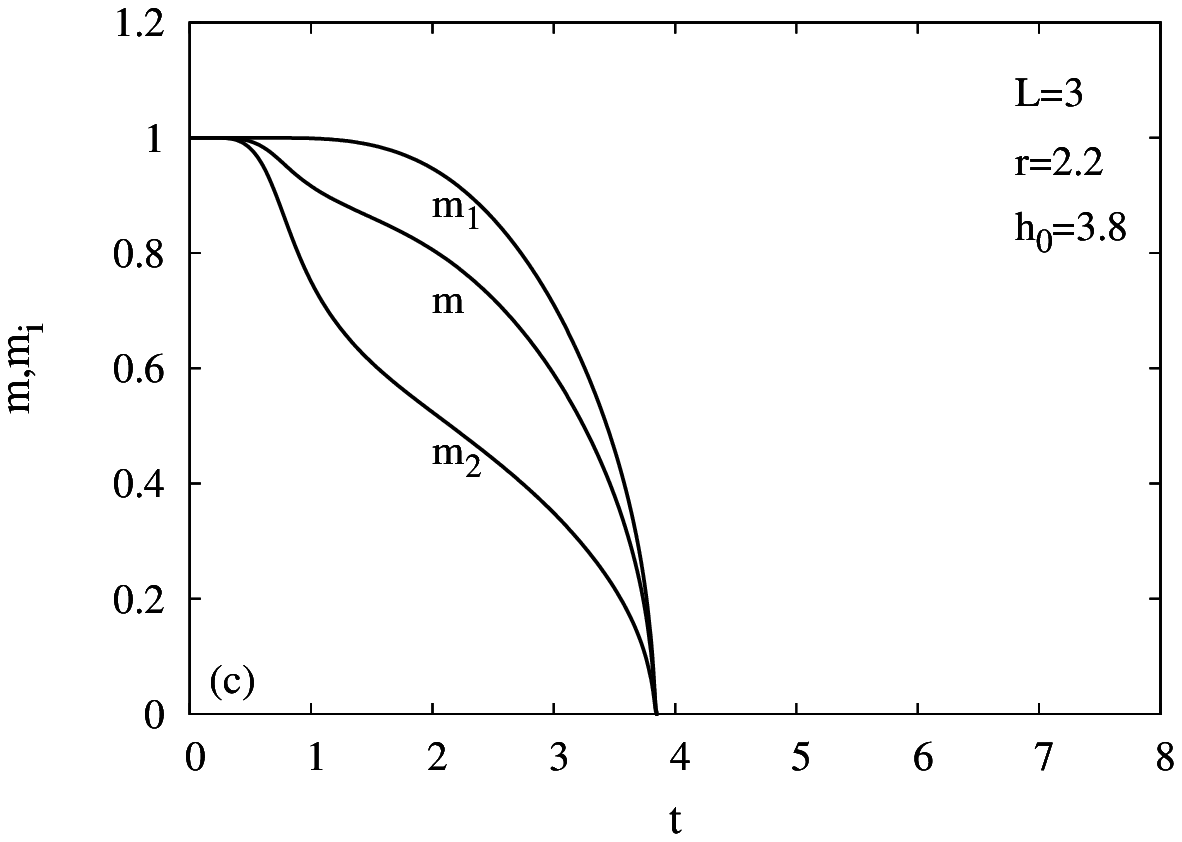, width=6cm}
\epsfig{file=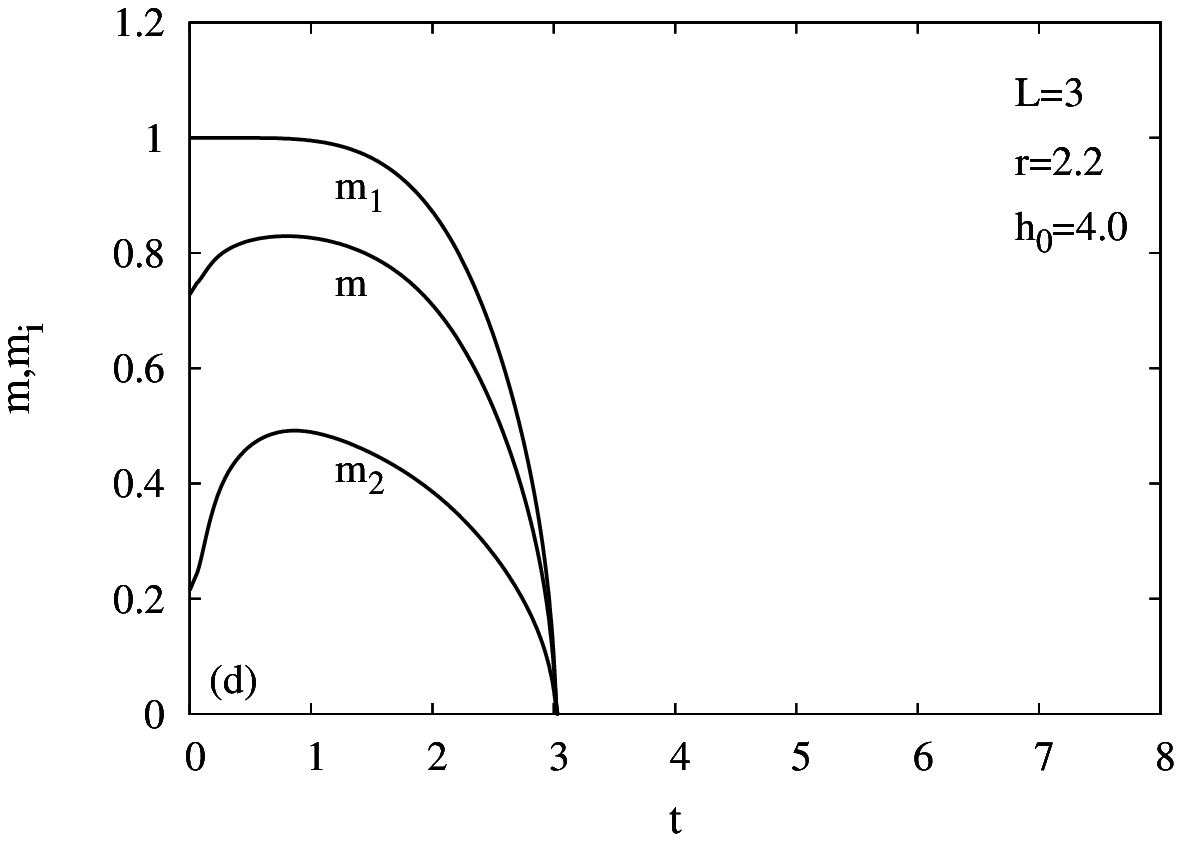, width=6cm}
\end{center}
\caption{Variation of the magnetization of the surface layer ($m_1$), inner layer ($m_2$), as well as the total magnetization ($m$) of the thin film with thickness $L=3$, with the temperature, for some selected values of $h_0$.  $r$ value is fixed as $r=2.2>r^*$. } \label{sek6}\end{figure}

\begin{figure}[h]\begin{center}
\epsfig{file=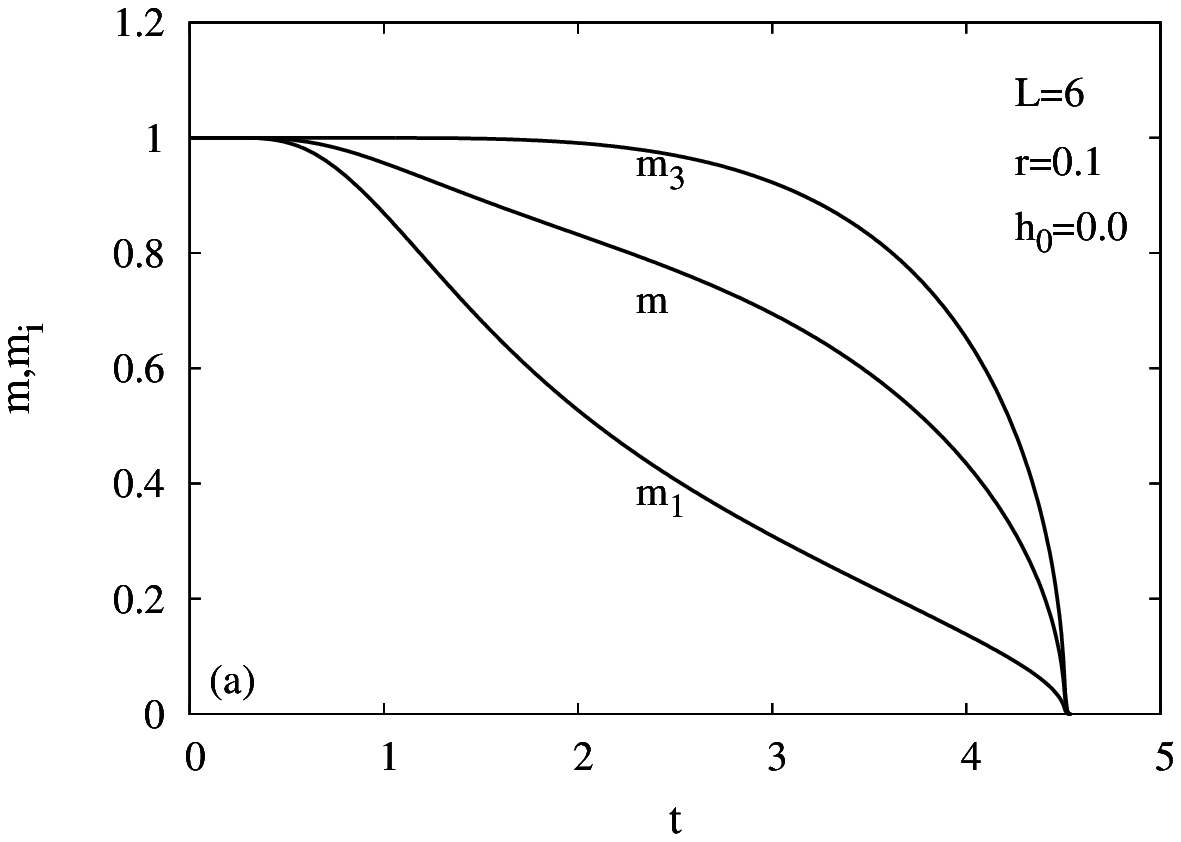, width=6cm}
\epsfig{file=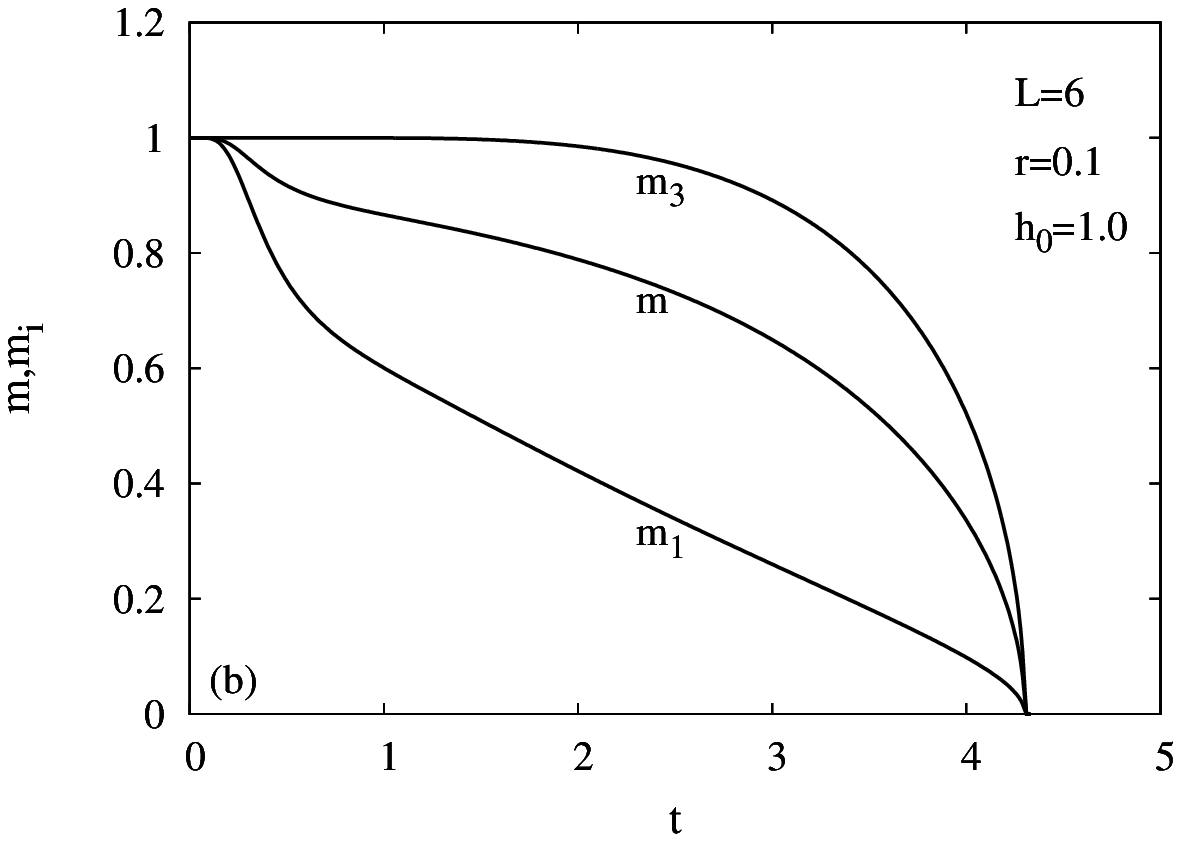, width=6cm}
\epsfig{file=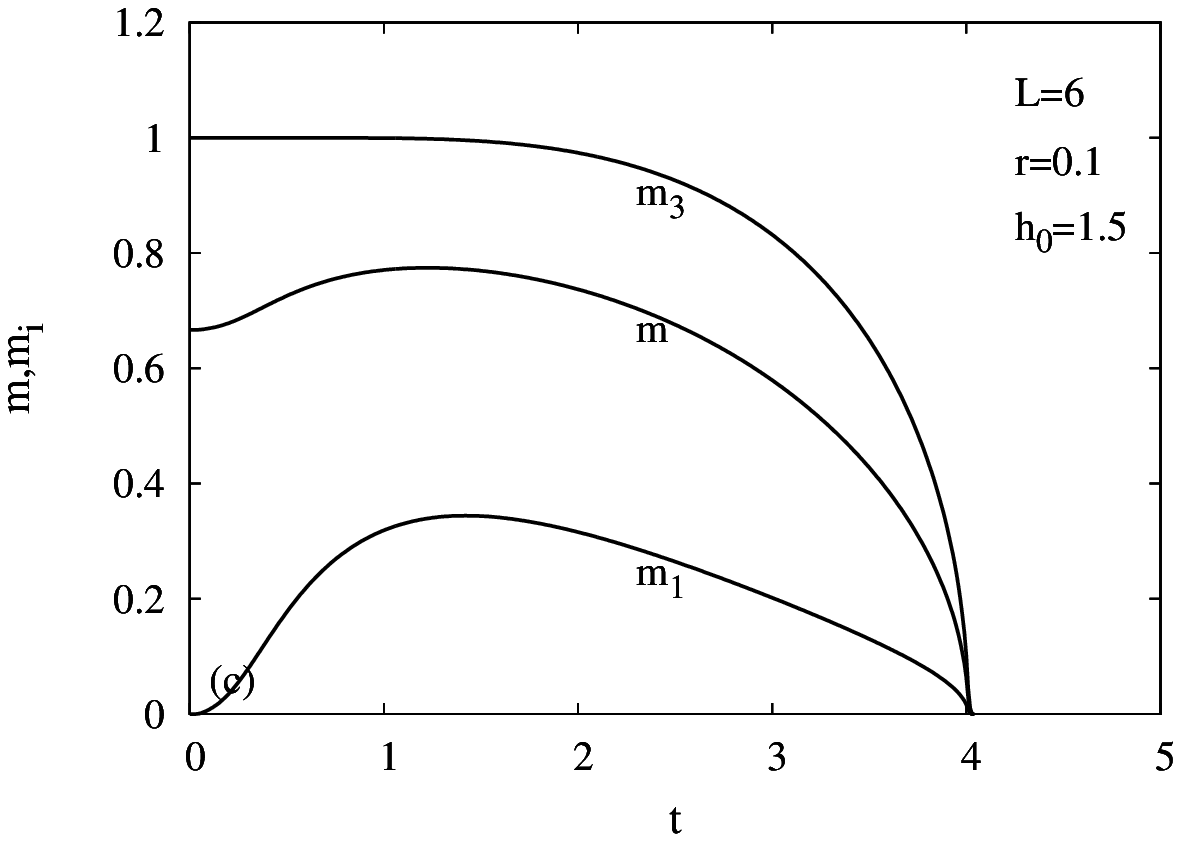, width=6cm}
\epsfig{file=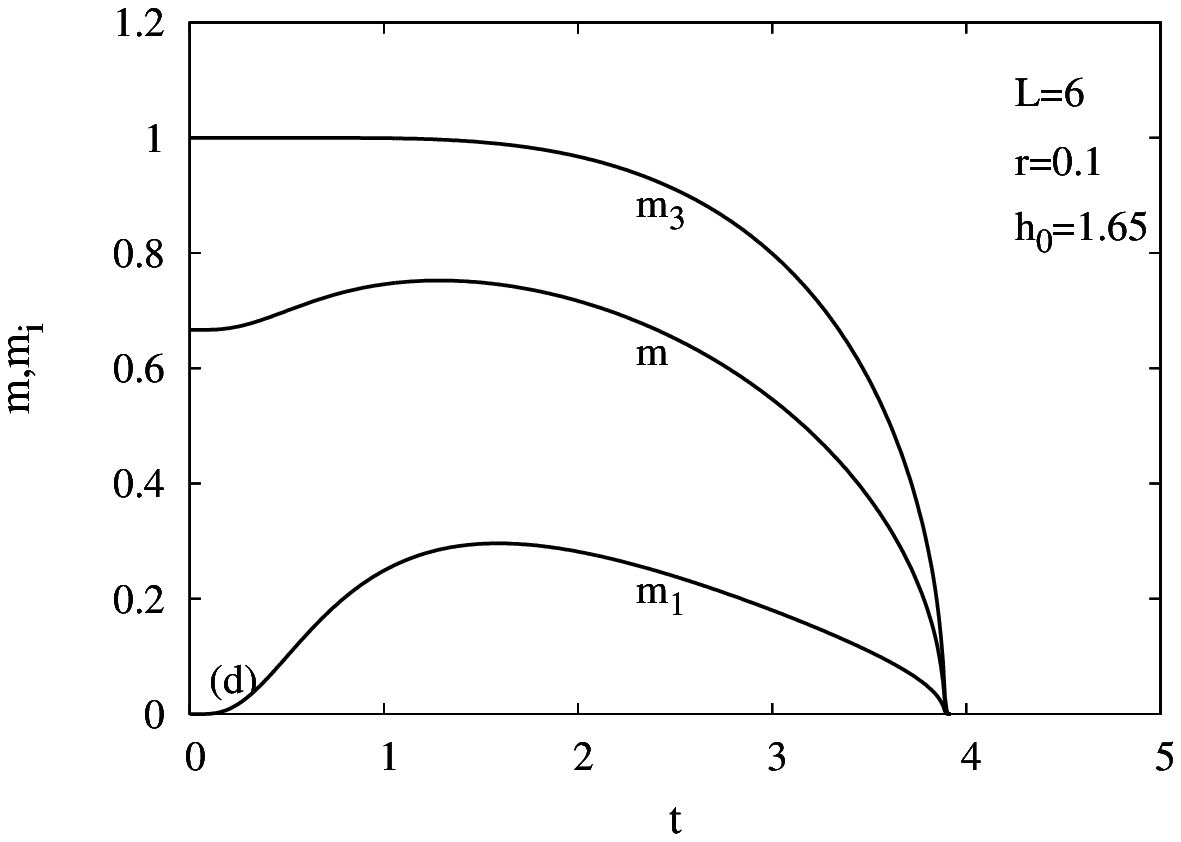, width=6cm}
\end{center}
\caption{Variation of the magnetization of the surface layer ($m_1$), inner layer ($m_3$), as well as the total magnetization ($m$) of the thin film with thickness $L=6$, with the temperature, for some selected values of $h_0$.  $r$ value is fixed as $r=0.1<r^*$.} \label{sek7}\end{figure}

\begin{figure}[h]\begin{center}
\epsfig{file=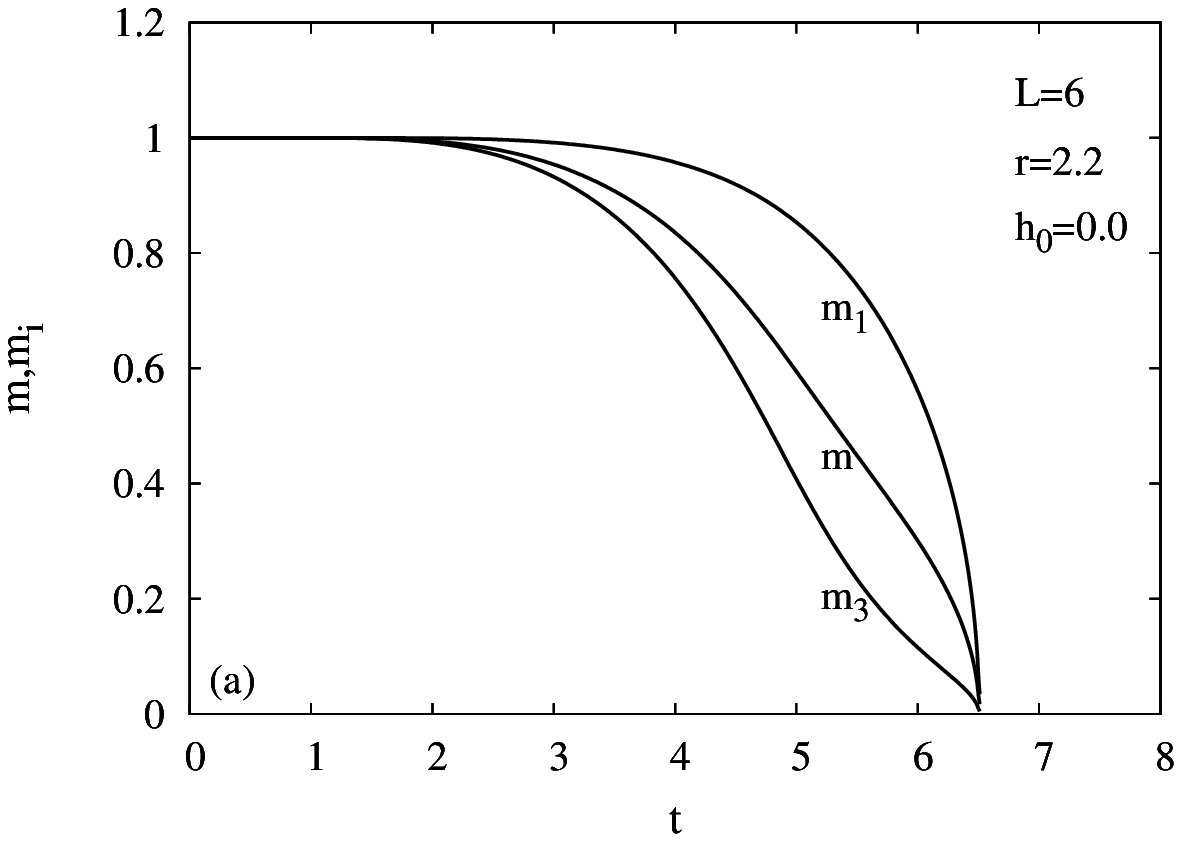, width=6cm}
\epsfig{file=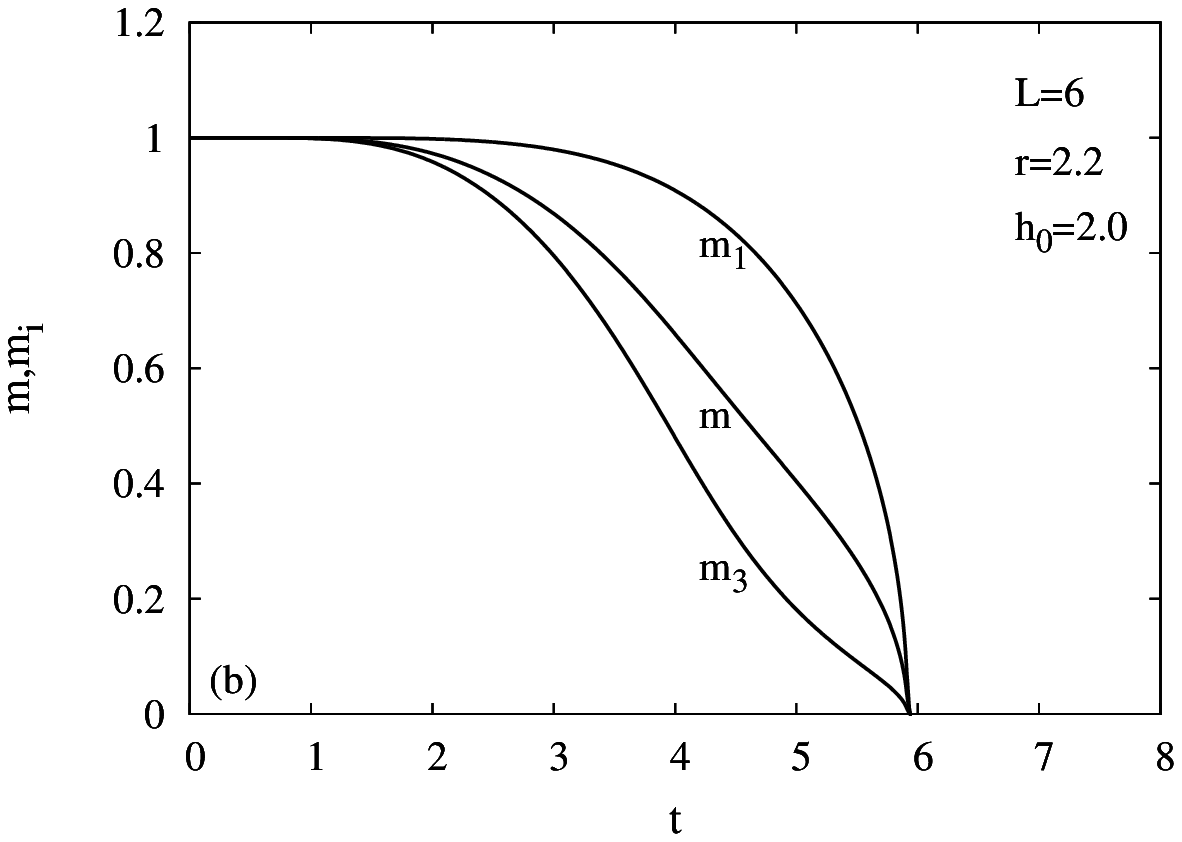, width=6cm}
\epsfig{file=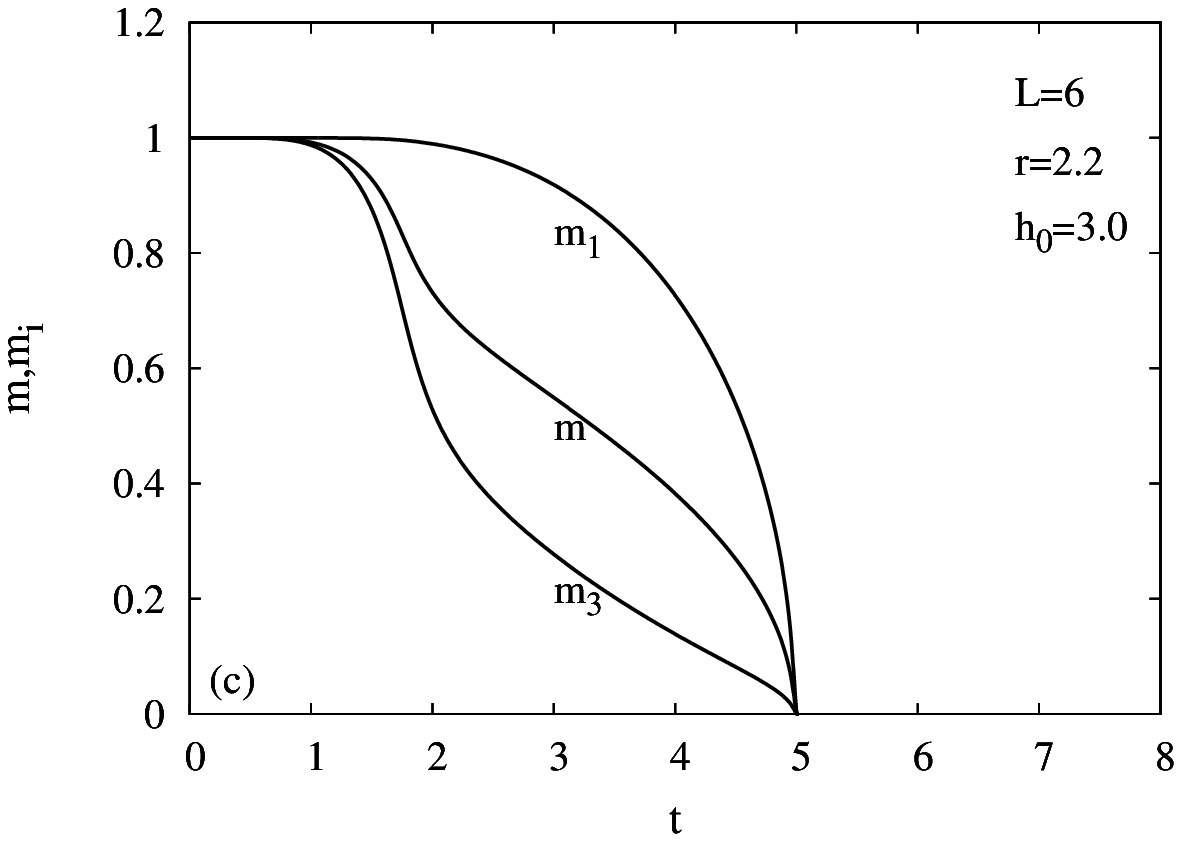, width=6cm}
\epsfig{file=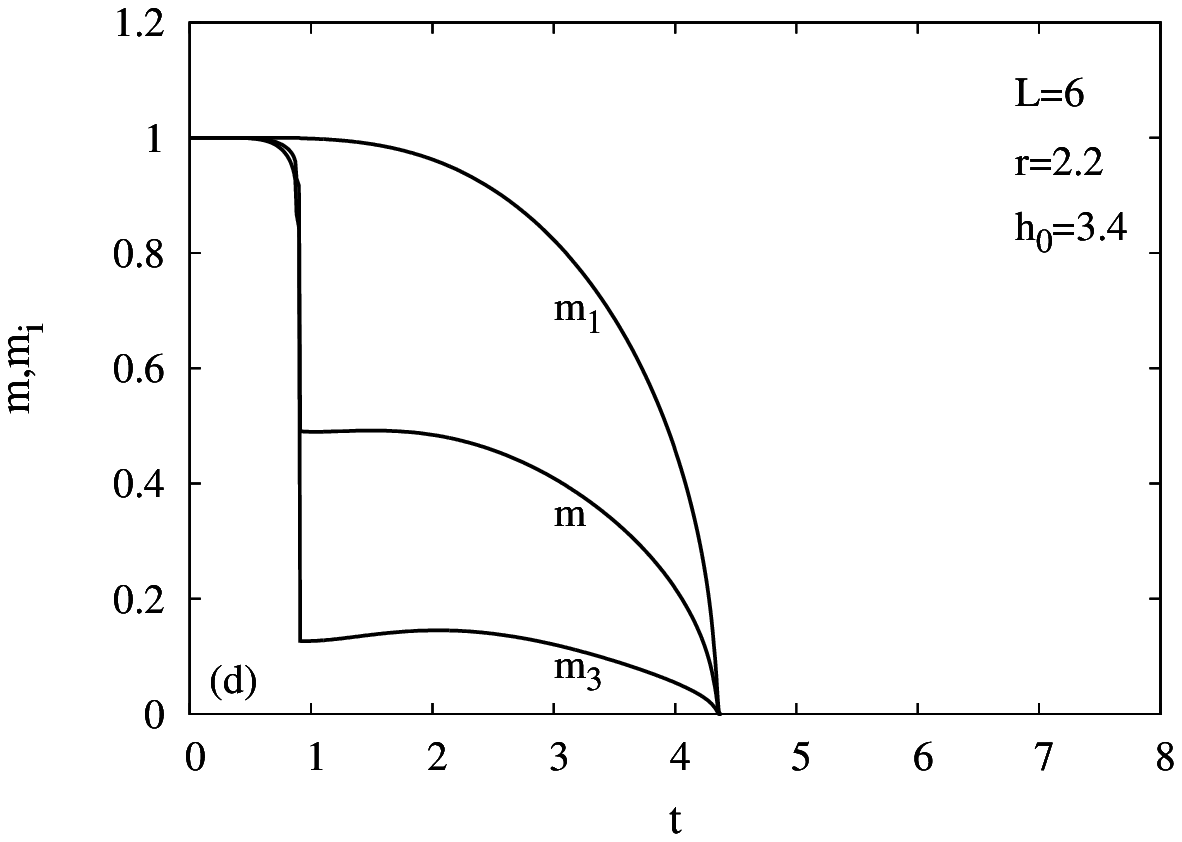, width=6cm}
\end{center}
\caption{Variation of the magnetization of the surface layer ($m_1$), inner layer ($m_3$), as well as the total magnetization ($m$) of the thin film with thickness $L=6$, with the temperature, for some selected values of $h_0$.  $r$ value is fixed as $r=2.2>r^*$.} \label{sek8}\end{figure}

As seen in Figs. \re{sek5} and \re{sek7} that, for the value of $r=0.1<r^{*}$, the magnetization of the surface layer has lower value than the magnetization of the inner layer,  both for $L=3$ and $L=6$. The reverse relation holds for the case $r=2.2>r^{*}$, can be seen in Figs. \re{sek6} and \re{sek8}.  Let us see for the case $r=0.1<r^{*}$ more closely. Both of the films that have thickness $L=3$ and $L=6$  completely ordered at the ground state for the magnetic field values $h_0=0.0$ and $h=1.0$ (see Figs. \re{sek5} and \re{sek7} (a),(b)). When the magnetic field rises this completely ordered state for the surface layer starts do destroyed. It can be seen from the Figs. \re{sek5} and \re{sek7} (c),(d) that, surface magnetization suppressed by the magnetic field distribution to the value of zero. For the $L=3$, this suppression
causes destroy of the ground state order of the inner layer partially (see. curves labeled by $m_2$ in Figs. \re{sek5} (c) and (d)). But the same situation is not available for the film that have thickness $L=6$ (see. curves labeled by $m_3$ in Figs. \re{sek7} (c) and (d)). This means that, when the film thickness rises, the effect of the completely disordered surface cannot penetrate the inner portions of the film. The same effect can be seen in the case  $r=2.2>r^{*}$, but this time for higher values of the magnetic field (see Figs. \re{sek6} and \re{sek8} (d)). We note that, in this case the magnetization of the surface layer has greater value than the magnetization of the inner layer, as mentioned above. All these facts show that, rising film thickness makes difficult to penetrate the randomness effects from the surface to the inner layers.

\section{Conclusion}\label{conclusion}

In this work, the effect of the bimodal random field distribution on the critical behavior of the isoropic Heisenberg thin films investigated.
As a formulation,  EFT-2 formulation has been used.

As in the bulk counterparts, rising randomness causes to decline of the critical temperature. Again to the similar results for the bulk system, tricritical behavior observed for the higher values of the center of the random field distribution $\pm h_0$. Rising randomness can induce first order transitions, regardless of the film thickness. This fact is shown also in magnetization-temperature behaviors. Besides, special point which equate all critical temperatures of the films that have different thickness decline, when $h_0$ rises. This fact is shown on the phase diagrams in the $(t_c-r)$ plane. It has been shown that, although the special point is not present, phase diagrams in that plane still exist. On the other hand similar trend has been obtained for the tricritical behavior. This behavior can appear after a specific value of $h_0$, then it disappears after a certain value of $h_0$. These two specific values are depend on the film thickness.

For the magnetic properties of the film, surface magnetization can lie below or above the magnetization of the inner layers. This situation depends on the value of $r$. When $r<r^*$  surface magnetization has smaller value than the magnetization of the inner layer. In this case it is observed that, rising randomness can induce completely disordered surface at zero temperature, while inner layer of the film can be ordered.  When the magnetization of the surface layer  depressed to zero at low temperatures, due to rising thermal fluctuation which comes from the rising temperature can create the non zero magnetization for the surface.

We hope that the results  obtained in this work may be beneficial form both theoretical and experimental point of view.

\end{document}